\documentclass[structabstract
]{aa}  
\usepackage[pdftex]{graphicx}
\usepackage{epstopdf}
\usepackage{natbib}
\usepackage{txfonts}
\usepackage{dcolumn}

\begin{document}
   \title{Multi-wavelength modeling of the spatially resolved debris disk of HD\,107146}
   \titlerunning{Multi-wavelength modeling of the HD\,107146 debris disk}

   \author{Steve Ertel\inst{1}
        \and
          Sebastian Wolf\inst{1}
        \and
          Stanimir Metchev\inst{2}
        \and
          Glenn Schneider\inst{3}
        \and
          John M. Carpenter\inst{4}
        \and
          Michael R. Meyer\inst{5}
        \and
          Lynne A. Hillenbrand\inst{4}
        \and
          Murray D. Silverstone\inst{6}
   }
   \institute{Institut f\"ur Theoretische Physik und Astrophysik, Christian-Albrechts-Universit\"at zu Kiel, Leibnizstra{\ss}e 15, 24098 Kiel, Germany \\
          \email{sertel@astrophysik.uni-kiel.de} 
        \and
          Department of Physics and Astronomy, State University of New York, Stony Brook, NY 11794-3800, USA
        \and
          Steward Observatory, University of Arizona, 933 North Cherry Avenue, Tucson, AZ 85721, USA
        \and
          Department of Astronomy, California Institute of Technology, 1200 East California Boulevard, Pasadena, CA 91125, USA
        \and
          Institute for Astronomy, ETH Z\"urich, Wolfgang-Pauli-Stra{\ss}e 27, 8093 Z\"urich, Switzerland
        \and
          Department of Physics \& Astronomy, University of Alabama, 326 Gallalee, Tuscaloosa, AL 35487-0324, USA
   }
   \date{}

 
  \abstract
   {}
   {We aim to constrain the location, composition, and dynamical state of planetesimal populations and dust around the young, sun-like (G2\,V) star \object{HD\,107146}.}
   {We consider coronagraphic observations obtained with the Advanced Camera for Surveys ({\it HST}/ACS) onboard the {\it HST} \normalfont in broad V ($\lambda_{c}\approx 0.6\,\mu\rm{m}$) and broad I ($\lambda_{c}\approx 0.8\,\mu\rm{m}$) filters, a resolved 1.3\,mm map obtained with the {\it Combined Array for Research in Millimeter-Wave Astronomy} \normalfont ({\it CARMA})\normalfont, {\it Spitzer}/IRS low resolution spectra in the range of $7.6\,\mu\rm{m}$ to $37.0\,\mu\rm{m}$, and the spectral energy distribution (SED) of the object at wavelengths ranging from $3.5\,\mu\rm{m}$ to $3.1\,\rm{mm}$. We complement these data with new coronagraphic high resolution observations of the debris disk using the Near Infrared Camera and Multi-Object Spectrometer (\itshape{HST}\normalfont/NICMOS) aboard the \itshape{Hubble Space Telescope} \normalfont (\itshape{HST}\normalfont) in the F110W filter ($\lambda_{c}\approx 1.1\,\mu\rm{m}$). The SED and images of the disk in scattered light as well as in thermal reemission are combined in our modeling using a parameterized model for the disk density distribution and optical properties of the dust.}
   {A detailed analytical model of the debris disk around HD\,107146 is presented that allows us to reproduce the almost entire set of spatially resolved and unresolved multi-wavelength observations. Considering the variety of complementary observational data, we are able to break the degeneracies produced by modeling SED data alone. We find the disk to be an extended ring with a peak surface density at 131\,AU. Furthermore, we find evidence for an additional, inner disk probably composed of small grains released at the inner edge of the outer disk and moving inwards due to Poynting-Robertson drag. A birth ring scenario (i.e., a more or less broad ring of planetesimals creating the dust disk trough collisions) is found to be the most likely explanation of the ringlike shape of the disk.}
   {}


   \maketitle

\section{Introduction}
\label{intro}

Debris disks were first discovered via infrared excess emission associated with main sequence stars such as Vega from observations with the {\it Infrared Astronomical Satellite} \citep[{\it IRAS};][]{aum84}. The first spatially resolved image of such an object was obtained around $\beta$\,Pictoris \citep{smi84}. In the last few years the {\it Spitzer Space Telescope} has increased the number of known debris disk candidates identified by their thermal emission above photospheric levels to several hundreds. As the most readily detectable signposts of other planetary systems, debris disks help us to improve our understanding of the formation and evolution of them as well as of our own solar system \citep{mey07, wya08, kri10}. The dust detected in such disks is thought to be removed from those systems by the stellar radiation on short time scales compared to their ages. Thus, it must be transient, or more likely continuously replenished by ongoing collisions of bigger objects like planetesimals left over from the planet formation process. For isolated systems the circumstellar dust's spatial distribution can be influenced by gravitational interaction with planets in addition to the influence of the star \citep{dom03, ken04, str06, wya08}. However, studying the spectral energy distribution (SED) of the dust alone provides only weak, ambiguous constraints for its properties such as chemical composition, grain size, and spatial distribution \citep{wol03}. For example, the location of the inner disk radius of the dust distribution is strongly degenerate with the lower limit of the grain size distribution. Direct measurements of the spatially resolved surface brightness by imaging mitigate many of these degeneracies. Images of disks seen near to face-on are particularly useful to directly determine the azimuthal and radial distribution of the dust and thus to trace the density structure.

\begin{figure*}
\centering
\includegraphics[width=1\linewidth]{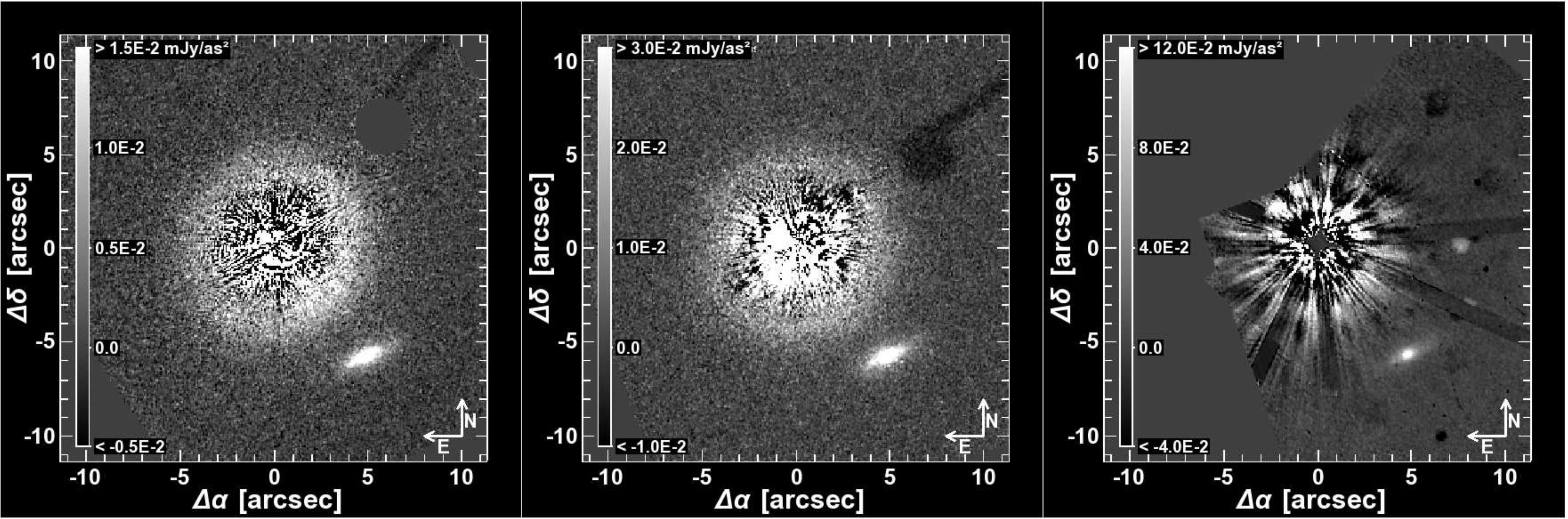}
\caption{Scattered light images. The {\it HST}/ACS images at $0.6\,\mu\rm{m}$ (left) and at $0.8\,\mu\rm{m}$ (center) and the {\it HST}/NICMOS image at $1.1\,\mu\rm{m}$ (right). The images are oriented north-up and the {\it HST}/ACS images are rebinned to the {\it HST}/NICMOS scale of $0.076\,\rm{\arcsec/pix}$. The plotted coordinate system indicates the distance from the star in arcseconds along the two axes. Note the different flux scales in $\rm{mJy/as}^{2}$ and the fact that the stellar PSF residuals are heavily truncated in these scales. The object in the SW is a background galaxy.}
\label{imaSca}%
\end{figure*}

In this paper, we present new spatially resolved observations of the HD\,107146 debris disk obtained with the Near Infrared Camera and Multi-Object Spectrometer (NICMOS) onboard the {\it Hubble Space Telescope} ({\it HST}). These data complement existing spatially resolved images at optical and millimeter wavelengths as well as spatially unresolved data of the star and the disk over a broad range of wavelengths (Table~\ref{SED_data}). An approach to combine all available data and so to derive a detailed model of the disk and to break degeneracies produced by modeling single datasets is presented.

HD\,107146 is a young, well studied, solar-type star (G2\,V), $28.5\pm0.7\,\rm{pc}$ \citep{per97} from the sun. Its luminosity and effective temperature are $\approx 1.1\,\rm{L_{\sun}}$ \citep{ard04} and $\approx 5850\,\rm{K}$, respectively. Its estimated age ranges from 80 to 320\,Myr \citep{wil04, moo06, roc09}. \citet{sil00} was the first to note the presence of an IR excess based on {\it IRAS} measurements. The origin of this excess in a debris disk system was discussed by \citet{met04}. HD\,107146 has been well studied in the context of the {\it Spitzer} Legacy Science Program {\it FEPS} \citep[Formation and Evolution of Planetary Systems;][]{mey06} resulting in a broad variety of photometric measurements as well as spectra of the star and the disk \citep[e.g.][]{hil08}. The debris disk around HD\,107146 is remarkable in many respects. With its large inner hole of $\sim 60\,\rm{AU}$ and its extension of $> 200\,\rm{AU}$ in radius, corresponding to $2\arcsec$ and $7\arcsec$, respectively \citep{ard04}, it can be imaged by coronagraphy without masking large parts of the disk. Its particularly high scattered light surface brightness enables high signal-to-noise imaging \citep{ard04}. The disk is seen with an inclination of $25^\circ \pm 5^\circ$ from face-on \citep{ard04}. This means, that the radial surface brightness distribution can be observed without strong projection effects which would occur in a close to edge-on case. With the inclination known, one can properly account for the remaining effects. In addition, the inclination of the disk enables us to constrain the dust properties from potential scattering phase asymmetries.

We present our new observations in Sect.~\ref{currObs}. The available data collected from the literature is summarised in Sect.~\ref{avData}. Details about our modeling, particularly the modeling software, the applied model, the order in which we include the available data and the information we extract in each step, respectively, can be found in Sect.~\ref{modeling}. The results are presented and discussed in detail in Sect.~\ref{discussion}. Conclusions are given in Sect.~\ref{conc}.

\section{New observations}
\label{currObs}

{\it HST}/NICMOS coronagraphic observations of HD\,107146 were obtained on 24 January  2005 as part of a larger circumstellar disk imaging survey\footnote{http://nicmosis.as.arizona.edu:8000/POSTERS/AAS\_JAN\_2006\\\_GO10177.pdf} ({\it HST} GO/10177; PI: Schneider) designed to reveal and spatially map near-infrared starlight scattered by dust orbiting the central stars in both protoplanetary and more evolved debris systems \citep{schn05}. Our HD\,107146 observations were conducted in our pre-planned survey mode, unchanged as a result of the discovery of a circumstellar scattered light disk around HD\,107146 at $0.6\,\mu\rm{m}$ and $0.8\,\mu\rm{m}$ wavelengths with ACS coronagraphy \citep{ard04} shortly before our already committed NICMOS observations. 

The HD\,107146 field was observed with NICMOS camera 2 in two contiguous {\it HST} visits at celestial orientations differing by 29\fdg9 in a single $54\,\rm{min}$  {\it HST} target visibility interval. HD\,107146 was placed behind the 0\farcs3 radius coronagraphic hole using ACCUM mode target acquisition exposures ($0.75\,\rm{s}$) in the F187N filter at each visit. Three $224\,\rm{s}$ (STEP32/NSAMP14) multiaccum exposures were taken in the F110W ($\lambda_{\rm{eff}} = 1.10\,\mu\rm{m}$, $\rm{FWHM} = 0.59\,\mu\rm{m}$) filter. Following the deep coronagraphic images, short pairs of $8.53\,\rm{s}$ (SCAMRR/NSAMP21) F110W exposures of the star were obtained with the star moved outside of the coronagraphic hole to directly enable flux density scaling with template point spread function (PSF) candidate stars similarly observed in GO/10177.  In each visit, the individual instrumentally calibrated count-rate images derived from the coronagraphic multiaccum exposures, were median combined and corrected for the well characterized 0.7\,\% geometrical distortion in NICMOS camera 2 (rectified to a two dimensional linear image scale of $75.8\,\rm{mas}\,\rm{pix}^{-1}$) yielding two images at different celestial orientations each with an effective integration time of $672\,\rm{s}$. Additional technical details concerning the observational strategy and data reduction techniques employed for all GO/10177 targets are presented in \citet{schn05, schn06}.

In our survey, in addition to our science targets, we similarly (but more deeply) observed five very bright stars of spectral types B through M (spanning the range of our targets) that were qualified through previous {\it HST} observations for use as coronagraphic PSF subtraction templates\footnote{see: http://www.stsci.edu/observing/phase2-public/10177.pro}, including GJ\,517 (K5V; J=6.876) and GJ\,355 (K0V; J=6.076).  The set of five coronagraphic PSF images created from these observations were co-aligned and flux-density scaled to HD\,107146 as described in \citet{schn05}. Circumstellar scattered light excesses were seen  at $r >$ 2\arcsec\ in both fully reduced coronagraphic images of HD\,107146 after separately subtracting each of the candidate PSF template images. These circumstellar structures, which conform to the morphology of HD\,107146 seen in scattered light at optical wavelengths as reported by \citet{ard04} from ACS imagery, are absent in identically processed PSF minus PSF template subtractions (as shown, for example, in Fig.~1 of \citet{schn06} for two of the same PSF templates used here).

The detailed sub-structures within the GJ\,517 and GJ\,355 coronagraphic PSFs were most closely matched to those arising from the underlying stellar component in the HD\,107146 images, and when separately subtracted yielded the most artifact-free circumstellar disk images. This was not wholly unexpected since GJ 517 and GJ 355 are closer in $|\rm{J-H}|$ color to HD\,107146 than our other PSF candidate stars\footnote{All three of these stars were observed after a first in four year mid {\it HST} Cycle 13 focus/desorption compensation move of the HST secondary mirror that additionally constrained PSF template suitability in our GO/10177 targets.}. While GJ\,355 is nearly the same brightness as HD\,107146 (in the F110W passband), GJ\,517 is significantly fainter ($\Delta\rm{J} = 1.0$). This deficiency was partially mitigated by the longer and deeper integrations employed in our F110W PSF template star observations (for GJ\,517, $1120\,\rm{s}$ total integration time) compared to our disk targets.  

The four PSF subtracted images of HD\,107146 using GJ\,517 and GJ\,355 as PSF subtraction templates were median combined, after rotating all about the co-aligned position of the occulted star to a common celestial orientation, except in individually masked and rejected regions significantly degraded by post-subtraction residual optical artifacts (primarily, insufficiently suppressed diffraction spikes from the HST secondary mirror support). This resulting two-orientation re-combined image that reveals the HD\,107146 circumstellar scattered light disk is shown in Fig.~\ref{imaSca} along with the ACS images (Sect.~\ref{avData}) that are used in the scattered-light analysis discussed subsequently in this paper. 

Because of the nature of the survey observations (with modal overheads for two-roll coronagraphy) allocating only a single spacecraft orbit per disk target, the NICMOS integration was rather shallow (22 minutes, combining all exposures at both field orientations) compared to ACS imaging in GO program 10330 (H. Ford PI) at shorter wavelength ($\approx$ 39\,min and 36\,min at 0.6 and $0.8\,\mu\rm{m}$, respectively). The $1.1\,\mu\rm{m}$ two-roll re-combined PSF subtracted image of the HD 107146 circumstellar disk (Fig.~\ref{imaSca}, right panel) is shown in a bi-polar linear display stretch in surface brightness from $-40\,\mu\rm{Jy/as}^{2}$ (hard black) to $+120\,\mu\rm{Jy/as}^{2}$ (hard white) alongside the PSF subtracted images obtained with ACS as discussed by \citet{ard04}. The circumstellar disk seen in the NICMOS image is instrumentally truncated in a small sector to the northeast at the apparent radius of the bright debris ring of $\approx 3\farcs5$ due to the field-of-view limitations of NICMOS camera 2, but is clearly revealed at all other circumstellar azimuth angles. Residual artifacts after PSF subtraction (emphasized for visibility relative to the brightness of the debris ring in the image display for illustrative purposes with the stretch chosen) are manifest primarily in two forms. First, high spatial frequency radial "streaks" and, second, small amplitude zonal under (or over) subtractions at some radii (the mottled light and dark backgrounds at low spatial frequency), both which often arise in imperfect NICMOS coronagraphic PSF subtractions. Two "objects" appear in the FOV, a galaxy $\approx$ 6\arcsec\ from HD 107146 (to the SW) also seen in the ACS images, and a slightly diffused point-like image artifact (at $\Delta\alpha = +7\farcs8$, $\Delta\delta = 0.0\arcsec$) persisting from the preceding target acquisition imaging. Neither of the PSF subtraction artifacts significantly affect the visibility of the disk, but their appearance does somewhat degrade the fidelity of the image.

\section{Available data and previous work}
\label{avData}

In addition to the {\it HST}/NICMOS observations we consider various already published observations. In this section these data as well as some more theoretical works including HD\,107146 are described in chronological order.

\citet{sil00} identified HD\,107146 as a candidate for having a Vega-like disk based on {\it IRAS} excess at wavelengths of $60\,\mu\rm{m}$ and $100\,\mu\rm{m}$. \citet{met04} performed 10 micron observations of HD\,107146 using the Palomar 5\,m Telescope. While the $10\,\mu\rm{m}$ emission was found to be photospheric, they independently from \citet{sil00} identified HD\,107146 as a Vega-like candidate based on the already mentioned {\it IRAS} observations. They estimated a stellar age of 50 to 250\,Myr and a minimum dust mass of $6.6 \times10^{-7}\,\rm{M_{\sun}}$ assuming gray body grains. Different masses were also derived using blackbody grains ($6.0 \times10^{-9}\,\rm{M_{\sun}}$) and ISM-like grains ($6.0 \times10^{-6}\,\rm{M_{\sun}}$) illustrating the strong dependence of the derived mass on assumptions of grain composition.

\citet{wil04} performed \itshape{SCUBA} \normalfont observations of the HD\,107146 disk at $450\,\mu$m and $850\,\mu$m. They found the disk to be marginally resolved with an extent of $300\,\rm{AU} \times 210\,\rm{AU}$ ($10\farcs5 \times 7\farcs4$) at $450\,\mu$m ($\rm{beam~size} = 8\farcs0$) but to be unresolved at $850\,\mu\rm{m}$ ($\rm{beam~size} = 14\farcs5$). They also performed \itshape{Keck\,II} \normalfont observations using the facility instrument LWS at $11.7\,\mu\rm{m}$ and $17.8\,\mu\rm{m}$, but did not detect any extended structure. They also modeled the disk's SED using a single temperature modified black body and found clear evidence of a large inner hole ($\geqslant 31\,\rm{AU}$ in radius). Based on this modeling they estimated a dust mass of $\approx 3\times10^{-7}\,\rm{M_{\sun}}$. A detailed discussion of the age of the star (30 to 250\,Myr) was given in \citet{wil04}.

\citet{ard04, ard05} resolved the disk using \itshape{HST}\normalfont/ACS PSF subtracted coronagraphic observations in the F606W (broad V) and F814W (broad I) filters (Fig.~\ref{imaSca}). These data are the basis for our resolved modeling of the debris disk. They found the disk to be a broad ring with maximum vertical optical depth at 130\,AU from the star and a FWHM of 85\,AU. They fitted elliptical isophotes to the disk images and found the disk inclined $25^\circ \pm 5^\circ$ from a pole-on orientation with the disk minor axis at a celestial position angle (east of north) angle of $58^\circ \pm 5^\circ$. \citet{ard04} also pointed out that the debris disk exhibits a surface brightness asymmetry in their ACS data, consistent with forward scattering of the star light by micron-sized dust grains given the inferred disk inclination.

\citet{car05} marginally resolved the disk at 3\,mm using the {\it Owens Valley Radio Observatory} ({\it OVRO}) millimeter-wave array. They found the total flux density at 3\,mm to be consistent with the previous work of \citet{wil04}. The angular size of the disk in that image is consistent with that derived from all other data, but the structure seems to be much more elongated. Because of the low resolution and low signal to noise ratio (SNR) we do not use this image for detailed modeling, but include its total flux density in our SED analysis.

\citet{kri08} compared results derived in their work from a collisional approach using a birth ring scenario to the SED of HD\,107146. They found the SED to be reproduced by a parent belt producing the dust at $\approx 200\,\rm{AU}$, too large compared to the available, spatially resolved data. This discrepancy was discussed in the context of roughness of Mie calculations due to the assumption of homogeneous spheres that might be a rough estimate of the shape of the dust grains.

\citet{hil08} presented unresolved {\it Spitzer} measurements at 3.6, 4.5, 8.0, 13, 24, 33, 70, and $160\,\mu\rm{m}$ all of which are included in our SED analysis. The low resolution spectra in the range of 7.6 to $37.0\,\mu\rm{m}$ obtained with {\it Spitzer}/IRS and used in their work are available through the Spitzer archive. We obtained a low resolution spectrum over the whole range as a combination of the available SL1, LL1, LL2 and LL3 spectra and included it in our present study.

\citet{cor09} presented a resolved 1.3\,mm map, obtained using \itshape{CARMA} \normalfont. In this image a ring-like structure is clearly seen. In comparison to the scattered light images of \citet{ard04} they found a less extended disk with the ring-like structure concentrated at $\sim 97\,\rm{AU}$ and two possible peaks in surface brightness at about the same distance from the star. A marginally resolved image at $350\,\mu\rm{m}$ is also presented. The 1.3\,mm map is considered in our resolved modeling of the disk.

\citet{roc09} computed some basic properties of the HD\,107146 debris disk using its SED (see their paper for details). They found an inner radius of $10.2\,\rm{AU}$, a minimum grain size of $8.6\,\mu\rm{m}$ and a dust mass of $3.3 \times 10^{-7}\,\rm{M_{\sun}}$ as best-fit values from their modeling. The inner radius as well as the minimum grain size reported by \citet{roc09} are inconsistent with that expected from the resolved data and the blow out size of the system. These discrepancies have been discussed in the context of grain segregation. They state that the modeled SED data predominantly trace the position of large (up to millimeter size) grains in contrast to the smallest grains (near the blow out size) traced by the {\it HST} scattered light images.

The compiled photometric data is listed in Table~\ref{SED_data}.

\begin{table}
\caption{Photometric data of the HD\,107146 system}
\label{SED_data}      
\begin{center}
\begin{tabular}{D{/}{}{4.6} D{/}{}{4.6} D{/}{}{4.6} c}        
\hline\hline                 
  \rm{Wavel}/\rm{ength} & \rm{Flu}/\rm{x} & \rm{Uncertai}/\rm{nty}~\rm{(}1\sigma~\rm{)} & Reference \\    
  \rm{[}\mu/\rm{m]} & \rm{[m}/\rm{Jy]} & \rm{[m}/\rm{Jy]} & \\
\hline                        
  3/.6    & 1711/.3   & 36/.7   & 1 \\
  4/.5    & 1074/.8   & 24/.7   & 1 \\
  8/.0    & 384/.4    & 8/.2    & 1 \\
  10/.3   & 247/.0    & 22/.0   & 2 \\
  11/.7   & 175/.0    & 10/.0   & 3 \\
  13/.0   & 138/.9    & 8/.5    & 1 \\
  17/.8   & 85/.0     & 8/.0    & 3 \\
  24/.0   & 59/.8     & 2/.5    & 1 \\
  33/.0   & 86/.7     & 5/.7    & 1 \\
  60/.0   & 705/.0    & 56/.0   & 4 \\
  70/.0   & 669/.1    & 47/.8   & 1 \\
  100/.0  & 910/.0    & 155/.0  & 4 \\
  350/.0  & 319/.0    & 45/.0   & 5 \\
  450/.0  & 130/.0    & 12/.0   & 6 \\
  850/.0  & 20/.0     & 3/.2    & 6 \\
  1300/.0 & 10/.4     & 1/.5    & 5 \\
  3100/.0 & 1/.42     & 0/.25   & 7 \\
\hline                                   
\end{tabular}
\end{center}
\tablebib{(1) \citet{hil08}; (2) \citet{met04}; (3) \citet{wil04}; (4) \citet{moo06}; (5) \citet{cor09}; (6) \citet{naj05}; (7) \citet{car05}}
\end{table}

\section{Analysis}
\label{modeling}

For our analysis of the debris disk we compare simulated data from a dust reemission modeling suite to the available spectral energy distribution and multi-wavelength imaging data.

\subsection{Model description}
\label{model}

The simulations of the dust reemission SED are done using the software tool \texttt{DDS}\footnote{http://www1.astrophysik.uni-kiel.de/dds/} \citep{wol05}. For simulations of resolved images our newly created tool \texttt{debris} is used. It allows to simulate images of arbitrary optically thin dust configurations in thermal reemission as well as in scattered light using exact Mie theory. The dust properties derived from Mie scattering are thereby computed with our software-tool \texttt{miex} \citep{wol04} included directly in \texttt{debris}.

The stellar emission from HD\,107146 is estimated by fitting a \citet{kur69} model to the $0.4-2.5\,\mu\rm{m}$ stellar SED. The fit gives $T_{\rm{eff}} = 5924\,\rm{K}$ and $\log g = 4.48$. A distance of the star from the sun of 28.5\,pc is adopted for our simulations. The derived uncertainties on the model parameters represent only the formal errors from our fitting. They do not include uncertainties in the distance of the star. For our current simulations of the disk an analytical, rotationally symmetrical density distribution $n(r)$ is employed ($r$: distance from the star). We limit on a single component model (no multi-ring structure) to minimize the degeneracies of our modeling. There is no obvious evidence for a multi-ring structure at this point of the work. A possible extension of our model motivated by the shape of the {\it Spitzer}/IRS spectrum is discussed in Sect.~\ref{spitzer-fit}.

\subsubsection{Employed density distribution}
\label{density_dist}

The simplest approach for the radial dust distribution is a power-law distribution $n(r) \propto r^{-\alpha}$ from an inner radius $R_{\rm{in}}$ to an outer radius $R_{\rm{out}}$ of the disk. As shown in Fig.~\ref{surf_bright_mod_ACS}, the observed radial brightness profiles derived from the {\it HST}/ACS images exhibit a decrease with increasing distance to the star, similar to a power-law distribution only in the outermost parts of the disk, outwards of $\approx 130\,\rm{AU}$ from the star. Further inward the surface brightness profiles flatten significantly. This is inconsistent with an increasing particle density. The profile derived from the F606W image even decreases with decreasing distance from the star inwards of $\approx 100\,\rm{AU}$. Inwards of this distance the profile derived from the F606W image is more reliable then that derived from the F814W image (Sect.~\ref{HST_fitting}). In the following we motivate and describe the employed density distribution.

We search for a simple analytical function which closely reproduces the disk density profile using a minimum of free parameters. Such a distribution allows simultaneous fitting of all parameters and can easily be compared to any distribution found in other studies. The actual distribution will be better described by a large number of physical parameters, reflecting all physical processes being responsible for it (e.g. the distribution of the planetesimals producing the dust, the effect of radiation pressure and Poynting-Robertson drag on the dust, mass and orbits of possible planets influencing the distribution by gravitational interaction). If a good fit is achieved, our empirical approximation will at least be very similar to this distribution.

As described above, the observed surface brightness profile can be described by an increase followed by a decrease with increasing distance from the star. In addition, the position of the peak and the width of the distribution have to be parameterized. Not limiting ourselves on functions with (known) physical interpretation, we consider the following candidate density distributions, that might be able to reproduce the described behaviour:
\begin{itemize}
 \item A product of two power-laws, one with positive and one with negative exponent: In such a distribution there is a strong correlation between the single parameters, making the fitting complicated and potentially resulting in very extreme, not physically interpretable values for all parameters.
 \item A product of two exponential functions can be simplified to one single exponential function and does not allow to reproduce the observed behaviour.
 \item A product of a Gaussian distribution and a power-law: Due to the symmetrical Gaussian dominating the behaviour of the function, distributions with strong asymmetries around the peak distance can only be achieved by very extreme indexes of the power-law part. Moreover the parameters in this distribution are heavily correlated making fitting complicated.
 \item A product of a power-law and an exponential function (analogous to Planck's law): As shown below, the number of parameters and correlations between them can be reduced to a minimum for this distribution.
\end{itemize}

Based on the above discussion we decide to reject the first three candidate distributions and to adopt the last one which is most convenient for our purpose. From an ad hoc point of view the following distribution can be employed:
\begin{equation}
  n'(r)=\left\{\begin{array}{ll}
    0 & \forall \ r < s \ \vee \ r < r_{\rm{sub}} \\[2mm]
    \left(\frac{r - s}{r_{0}}\right)^{\alpha_1}\cdot \exp \left[-\alpha_2 \left(\frac{r - s}{r_{0}}\right)\right] & \forall \ r \geqslant s \ \wedge \ r \geqslant r_{\rm{sub}}
  \end{array}\right.
\end{equation}
In this equation the quantity $r_{\rm{sub}}$ is the sublimation radius of the dust, $r_{0}$ is a scale length stretching the distribution along $r$, $s$ is a characteristic length shifting the distribution along $r$, and $\alpha_1$ and $\alpha_2$ are parameters describing the slopes of the dust distribution in the inner and outer region, respectively. In this equation one can express the quantity $r_{0}$ through the peak radius $r_{\rm{p}}$ of the distribution using the following condition:
\begin{equation}
  \frac{dn'(r)}{dr}(r = r_{\rm p}) = 0\ ; \hspace{0.5cm} r_{\rm p} > s
\end{equation}
Defining a new density distribution $n(r)$ normalised to $n(r_{\rm p}) = 1$, one can eliminate the parameter $\alpha_2$:
\begin{equation}
  \label{dens_final}
  n(r) = \left\{\begin{array}{ll}
    0 & \forall \ r < s \ \vee \ r < r_{\rm{sub}} \\[2mm]
    \left(\frac{r - s}{r_{\rm p} - s}\right)^{\alpha_1}\cdot \exp{\left[\alpha_1\left(1-\frac{r - s}{r_{\rm p} - s}\right)\right]} & \forall \ r \geqslant s \ \wedge \ r \geqslant r_{\rm{sub}}
  \end{array}\right.
\end{equation}
In the resulting density distribution there are only three free parameters, $s$, $r_{\rm p}$ and $\alpha_1$, describing the inner increase and the outer decrease of the density as well as its width and peak position. It is important to note, that the number of free parameters is still the same as for the power-law distribution ($\alpha$, $R_{\rm{in}}$, and $R_{\rm{out}}$), despite an apparently much more complex shape of the distribution. Furthermore, only two parameters, $s$ and $\alpha_1$, are correlated (see discussion in Sect.~\ref{HST_fitting}), while even in the case of a simple power-law $\alpha$, $R_{\rm{in}}$, and $R_{\rm{out}}$ are strongly correlated because of the poor fit on the observed data. In addition, due to the particular shape of this distribution, there will be usually no need for an inner cut off $R_{\rm{in}}$, which is essential in the power-law distribution.

In debris disks we often observe an additional outer break in the surface brightness suggesting a break in the radial density distribution \citep[e.g.][]{kal06}. Such a break can be realized in the same way as for the case of a power-law distribution employing an outer radius $r_{\rm{out}}$ of the density distribution: $n(r > r_{\rm{out}}) = 0$. While an outer radius of the disk is usually needed to limit the extension of the disk for a power-law distribution, that falls to negligible low values very slowly, the exponential decay reaches such values much earlier. So if there is no real, physical break observed in the surface brightness profile (as it is the case for HD\,107146), the outermost radius will just be larger than the flux limited extent of the disk, and is only used to limit the extension of our simulated images. Thus, in this study we set $r_{\rm{out}}$ to 500\,AU. In contrast to the power-law distribution with the two cut-off radii our new distribution is continuous and continuously differentiable. Due to its analytical character it can easily be compared to any distribution found in other studies. 

Additionally, a constant opening angle of the disk of $10^\circ$ is employed to define the vertical extension of our disk. That means the disk height is increasing linearly with distance from the star. As long as the angle is small enough, this assumption is of lower importance for our modeling. This simplification is chosen because of the nearly face-on orientation of the disk and allows one to compute the surface density distribution $\Sigma(\rho)$ from the radial density distribution $n(r)$ very easily:
\begin{equation}
  \label{eq1}
  \Sigma(\rho) \propto n(r)\cdot r
\end{equation}
Here the quantity $\rho$ is the distance from the star projected on the midplane of the disk.

\subsubsection{Employed dust grain properties}
\label{sizedist}

The only available data that could inform about the mineralogy of the dust grains is the {\it Spitzer} spectrum. Since this spectrum does not exhibit any significant features, we employ for the grain composition astronomical silicate with a bulk density of $2.7\,\rm{g/cm^{3}}$ \citep{dra84, wei01}. The grain size distribution follows a power-law
\begin{equation}
  \label{size_distr}
  n(a) \propto a^{-\gamma}
\end{equation}
from a minimum grain size $a_{\rm{min}}$ to a maximum grain size $a_{\rm{max}}$. We assume the grain size distribution to be the same at all distances, so we can compute the abundance $n(r,a)$ of a certain grain size at a certain position in the disk through
\begin{equation}
  n(r,a) \propto n(r) \cdot n(a)
\end{equation}

This assumption is necessary to limit the complexity of our model. It is not fully justified, since the effects of radiation pressure as well as Poynting-Robertson drag are stronger for smaller grains (larger $\beta$ ratio, Fig.~\ref{betaratio}) which naturally leads to grain segregation. As the disk is expected to be collision dominated (Sect.~\ref{HST_fitting}), we assume that grain segregation has a minor effect on the distribution of grains of different size. This assumption will prove successful to reproduce most of the available data on HD\,107146.

In our grain size distribution $a_{\rm{min}}$ and $\gamma$ are free parameters but $a_{\rm{max}}$ is fixed to 1\,mm, since the influence of bigger grains on the thermal reemission as well as on the scattered light can be neglected for significantly steep size distributions. \emph{Therefore the dust masses derived in this work represent the mass in particles smaller than 1\,mm in radius only.}

\subsection{General guideline}
\label{guideline}

In order to efficiently break the modeling degeneracies one has to understand the information that can be extracted from the observations and how to combine the ancillary data. In the following we present the general guideline used to reduce the number of free parameters to a minimum in each step of the modeling. At the end of the modeling study each of our free parameters described above has been fitted. The described approach leads to a complete set of model parameters for one unified model of the disk, consistent as far as possible with all included data.

\emph{Radial dust distribution from high resolution images:} Having high resolution scattered light images of a radially symmetrical, optically thin debris disk seen near to face-on, the surface number density profile $\Sigma(\rho)$ of the disk is correlated with the face-on surface brightness profile $\varphi (\rho)$ as follows:

\begin{equation}
  \Sigma (\rho) \propto \rho^{2} \cdot \varphi (\rho) \cdot \Phi(90^\circ) 
\end{equation}

Because the scattering efficiency at a scattering angle of $90^\circ$, $\Phi$($90^\circ$), of the dust is unknown at this point, one can not derive the absolute value of the surface density. For detailed fitting one has to make reasonable assumptions on certain dust properties (in our case $a_{\rm{min}}$ and $\gamma$, see Sect.~\ref{HST_fitting} for the assumptions made in the present work). This is possible, because these assumptions primarily affect the total brightness of the disk and thus the resulting disk mass but not the shape of the disk. The radial density distribution $n(r)$ can be found using Eq.~\ref{eq1}. Due to the near to face-on orientation, the effect of the vertical structure of the dust distribution on the surface brightness distribution is negligible. Because scattered light images trace particularly the smallest grains, the validity of the derived density distribution for all grain sizes has to be verified. Additionally, for coronagraphic scattered light images the inner parts of the disk are hidden by the coronagraph or heavily affected by the residuals of stellar PSF subtraction. Hence the density distribution in the inner region remains uncertain. Both the validity of the distribution for larger grain sizes as well as in the inner region can be verified including spatially resolved observations at thermal reemission wavelengths, e.g. from mid-infrared to millimeter.

\emph{Dust properties from SED modeling:} Once the spatial distribution of the dust is constrained from modeling the scattered light images\footnote{Simplifying assumption: density distribution is the same for all dust species (size, composition; see Sect.~\ref{sizedist})} the main degeneracy in SED fitting is broken. One can now derive properties of the dust grains, in our case the lower grain size and the exponent of the grain size distribution, from fitting the SED. One then has to go back to the scattered light images and confirm the results for the density distribution using these dust properties instead of the assumptions made before. Eventually, this process requires several iterations.

\emph{A complete set of model parameters:} With the final set of model parameters one can obtain (almost) independent mass estimates from all images as well as from the SED. These masses should be consistent. In the case of a slightly inclined disk and scattered light data available one can further verify the fit and break potential remaining degeneracies quantifying how well the scattering asymmetry is reproduced by the model.

\subsection{Modeling the Hubble scattered light images}
\label{HST_fitting}

\begin{figure}
\centering
\includegraphics[angle=270,width=1\linewidth]{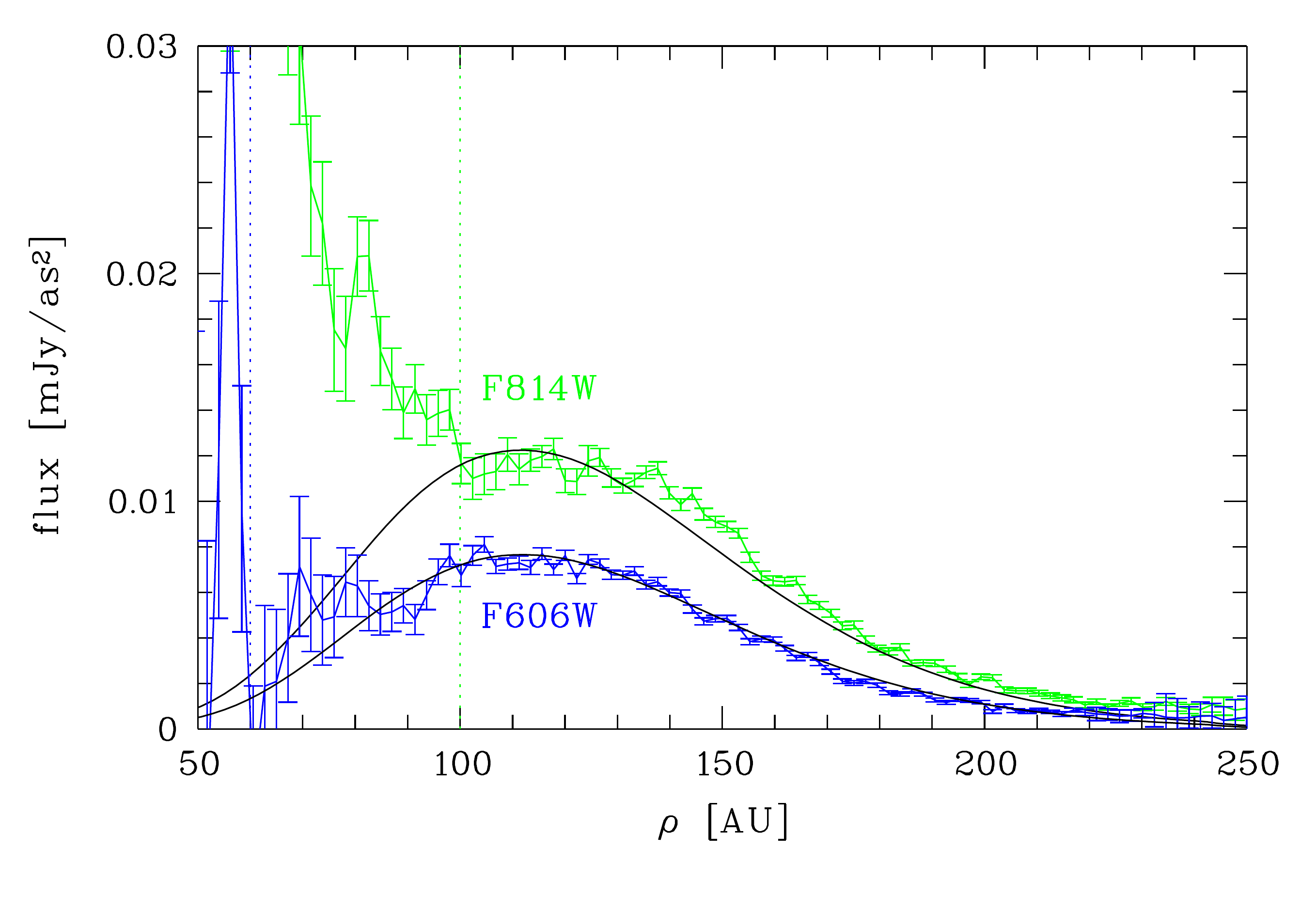}
\caption{Azimuthally medianed (one pixel wide annular zones) surface brightness profiles derived from the {\it HST}/ACS scattered light images after deprojection (for details see Sect. \ref{HST_fitting}). Note that the F606W radial brightness profile is reliable only for $\rho > 60\,\rm{AU}$ and the F814W image is reliable only for $\rho > 100\,\rm{AU}$ (dashed, vertical lines). An additional error of $\approx 5\%$ due to uncertainties in stellar PSF subtraction is not included in the plotted error bars. The stellar flux densities are $5.9 \pm 0.3\,\rm{Jy}$ and $6.9 \pm 0.3\,\rm{Jy}$ at the F606W and F814W wavebands \citep{ard04}. The black, solid lines represent the profiles derived from ouf final model of the disk (Sect.~\ref{modeling}). The albedo of our employed astronomical silicate grains (Sect.~\ref{sizedist}) is 0.56 and 0.57 at $\lambda = 0.6\,\mu\rm{m}$ and $\lambda = 0.8\,\mu\rm{m}$, respectively, for the smallest ($2.5\,\mu\rm{m}$) grains dominating the scattering.}
\label{surf_bright_mod_ACS}%
\end{figure}

For the two \itshape{HST}\normalfont/ACS images (Fig.~\ref{imaSca}) \citet{ard04} found that the artifacts from the PSF subtraction dominate the scattered light within $\approx 2\arcsec$ from the star. For a detailed discussion of the errors in these images see \citet{ard04}.

Fig.~\ref{surf_bright_mod_ACS} shows the sky-plane deprojected (i.e., if seen face-on and assuming $i = 25^\circ$) azimuthally medianed radial surface brightness profiles of the disk derived from the two {\it HST}/ACS images. They are obtained following \citet{wei99}. After deprojecting the images to face on, adopting the inclination of the disk of $25^\circ$ from \citet{ard04} and using flux conservation, the azimuthal median is derived in one pixel wide concentric radial zones. The error bar on each measurement is then computed following the equation
\begin{equation}
 \sigma = \frac{1}{N} \sqrt{\sum_{\rm i=1}^{N}\left(\bar{x} - x_{\rm i}\right)^{2}}
\end{equation}
where $N$ is the number of pixels in a radial zone, $\bar{x}$ is the median flux in a zone and $x_{\rm i}$ is the flux in one pixel. This is analogous to the standard error of the mean, but using the median in each zone. Deriving the azimuthal median does not account for the scattering asymmetries which are expected to be eliminated when medianing azimuthally. This assumption is less relevant because the same procedure has been applied deriving radial profiles from the modeled images, so measured and modeled profiles can be compared directly. In the F110W image large PSF subtraction residuals do not allow to derive a reliable profile.

The disk surface brightness derived from the two {\it HST/ACS} images decreases with increasing radius for $r > 130\,\rm{AU}$. A similar behaviour can also be seen qualitatively in the {\it HST}/NICMOS image. Within a radius of $130\,\rm{AU}$ of the star the two {\it HST}/ACS surface brightness profiles flatten significantly. Inwards of $\approx 100\,\rm{AU}$ the F814W image shows a behaviour that is completely different from that of the F606W image. While the F606W image shows a decrease in surface brightness with decreasing radius, the surface brightness obtained from the F814W image exhibits a steep increase. An examination of the F814W data shows that the best global scaling of the PSF template results in an under-subtraction of the stellar light at $2.0\arcsec \leqslant r \leqslant 3.5\arcsec$ ($57\,\rm{AU} \leqslant r \leqslant 100\,\rm{AU}$), particularly toward the SE. This becomes obvious when comparing the F814W and F606W images in Fig.~\ref{imaSca}. Furthermore, no physical process is known to us that could produce such a different behaviour at $0.8\,\mu\rm{m}$ compared to $0.6\,\mu\rm{m}$ and $1.1\,\mu\rm{m}$. Following the ACS Instrument Handbook, Version 8.0 \citep{bof07}, Sections 5.6.5, 6.2.9, and 9.3.2, and Version 10.0 \citep{may10}, Section 5.6.5, there is a known halo in the HRC images beyond $700\,\rm{nm}$. Furthermore, the PSF in this filter is relatively large, so that less light is blocked by the coronagraph. This results in a coronagraphic PSF that is more sensitive to spot shifts and star-to-spot misalignments. Thus, we attribute the increase of the surface brightness profile derived from the F814W image with decreasing distance from the star at $r \leqslant 100\,\rm{AU}$ to technical artifacts and residuals from imperfect PSF subtraction.

Due to the lower amplitude of the PSF subtraction residuals compared to the other scattered light images, the F606W image is the best choice for starting our simulations. To reduce the influence of PSF subtraction errors on our modeling, we mask the inner 2\arcsec\ ($\approx 60 AU$) in radius around the center of the image mentioned by \citet{ard04} and neglect it in the analysis. To obtain comparable results, the same region of our simulated images is masked. An error map for the PSF subtracted image is obtained from under- and over-subtractions of the PSF in each pixel derived from different PSF reference stars observed. The following equation is employed:
\begin{equation}
 \sigma_{i} = \frac{1}{2} \left(\,\left|\,\rm{hi}_{\it i} - \rm{low}_{\it i}\,\right|\,\right) + \rm{noise}
\end{equation}
In this equation $i$ runs over all pixels and $\rm{low}_{\it i}$ and $\rm{hi}_{\it i}$ are values obtained from the under- and over-subtraction (minimum and maximum value in one pixel subtracting the different PSF references). Additionally, we add the $1\sigma$ noise level derived from pixel to pixel noise in regions of the original image far from the star where no signal is detected.

\begin{figure}
\centering
  \includegraphics[angle=270,width=1\linewidth]{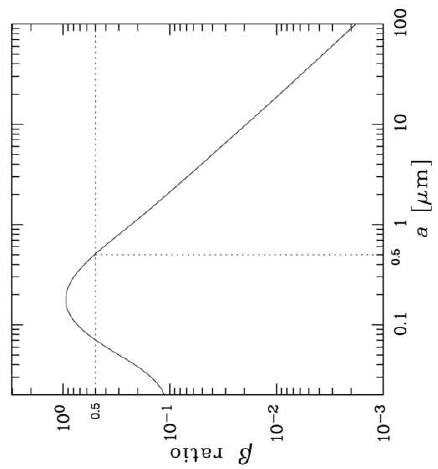}
\caption{$\beta$ ratio (the ratio between radiation pressure force and gravitational force) for astronomical silicate with a bulk density of 2.7\,g/cm$^3$ around HD\,107146. The dashed horizontal and vertical lines indicate $\beta = 0.5$ and the corresponding blow out size of $a = 0.5\,\mu\rm{m}$, respectively.}
\label{betaratio}
\end{figure}

To find the radial dust distribution, the parameters describing the grain size distribution are fixed and only the free parameters of our particle density distribution ($\alpha_1$, $r_{\rm{p}}$, and $s$) are varied. The parameters of the grain size distribution ($a_{\rm{min}}$, $a_{\rm{max}}$, and $\gamma$) are fixed as follows:
\begin{itemize}
 \item The value of $a_{\rm{max}}$ is fixed to 1\,mm (Sect.~\ref{sizedist}).
 \item The value of $a_{\rm{min}}$ is fixed at $1.0\,\mu\rm{m}$. The $\beta$ ratio, namely the ratio between radiation pressure force and gravitational force on a dust grain around HD\,107146 is computed (Fig.~\ref{betaratio}). A blow out size ($\beta = 0.5$) of $0.5\,\mu\rm{m}$ can be found. In contrast, the dust is scattering significantly red with respect to the stellar spectrum in the range of $0.6$ to $1.1\,\mu\rm{m}$ (Fig.~\ref{imaSca},~\ref{surf_bright_mod_ACS}), implying the absence of small, Rayleigh scattering dust grains. For the adopted dust properties, only a minimal grain size larger than $\approx 1.0\,\mu\rm{m}$ results in a red scattering disk in the observed range of wavelengths.
 \item A value of $\gamma = 3.5$ is chosen, being expected from an equilibrium collisional cascade \citep{doh69}. \citet{the07} modeled collisional processes in debris disks and found collision time scales $\tau_{\rm{coll}}$ much smaller than $10^{7}$ years for micron-sized particles, decreasing with increasing disk mass. The Poynting-Robertson timescale $\tau_{\rm{PR}}$ for a particle with $a_{\rm{min}} = 1\,\mu\rm{m}$ to spiral onto the star from 100 AU is calculated. A value of $1.7 \times 10^{7}$ years can be found. Taking into account the high mass of the disk \citep{wil04}, the expected collision time scale is about two orders of magnitude lower, e.g. $10^{5}$ years. Because of the decreasing influence of Poynting-Robertson drag for larger particles, we conclude that collisions dominate the evolution of the dust significantly ($\tau_{\rm{coll}}/\tau_{\rm{PR}} < 6.0 \times 10^{-3}$).
\end{itemize}

\begin{figure*}
\begin{minipage}{0.33\linewidth}
  \includegraphics[angle=270,width=1\linewidth]{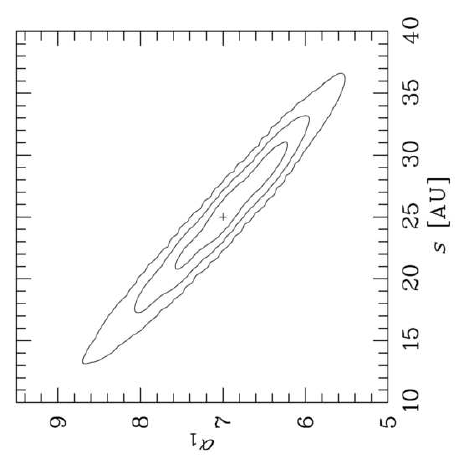}
\end{minipage}
\begin{minipage}{0.33\linewidth}
  \includegraphics[angle=270,width=1\linewidth]{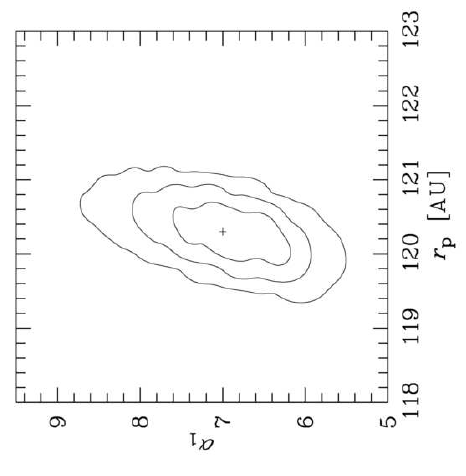}
\end{minipage}
\begin{minipage}{0.33\linewidth}
  \includegraphics[angle=270,width=1\linewidth]{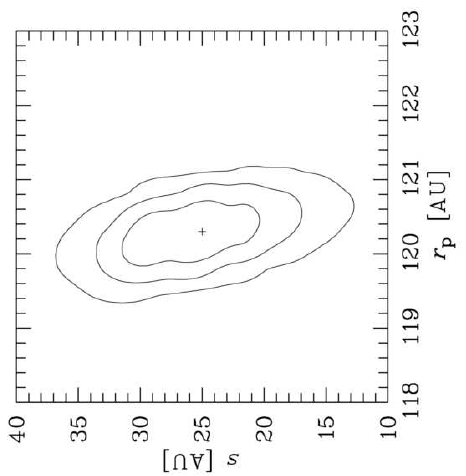}
\end{minipage}
\caption{Projection of the three dimensional ellipsoids of constant $\chi^{2}$ on the $r_{\rm{p}} = \rm{const}$ (left), $s = \rm{const}$ (center), and $\alpha_1 = \rm{const}$ (right) planes from fitting the F606W image. The contour levels show $1\sigma$, $2\sigma$, and $3\sigma$ levels of confidence, respectively. The cross indicates the position of the best-fit values. The plots illustrate very well the correlation between $\alpha_1$ and $s$ and the very weak correlations between these parameters and $r_{\rm{p}}$, respectively. For a detailed description of the parameters shown see Sect.~\ref{density_dist}.}
\label{map_paraspace}
\end{figure*}

For the adopted inclination of the disk of $25^\circ$ from \citet{ard04} the adopted values of $a_{\rm{min}}$, $a_{\rm{max}}$, and $\gamma$ produce an azimuthal scattering phase asymmetry in the surface brightness that is in good agreement with the observed one.

\begin{figure*}
\centering
\includegraphics[width=1\linewidth]{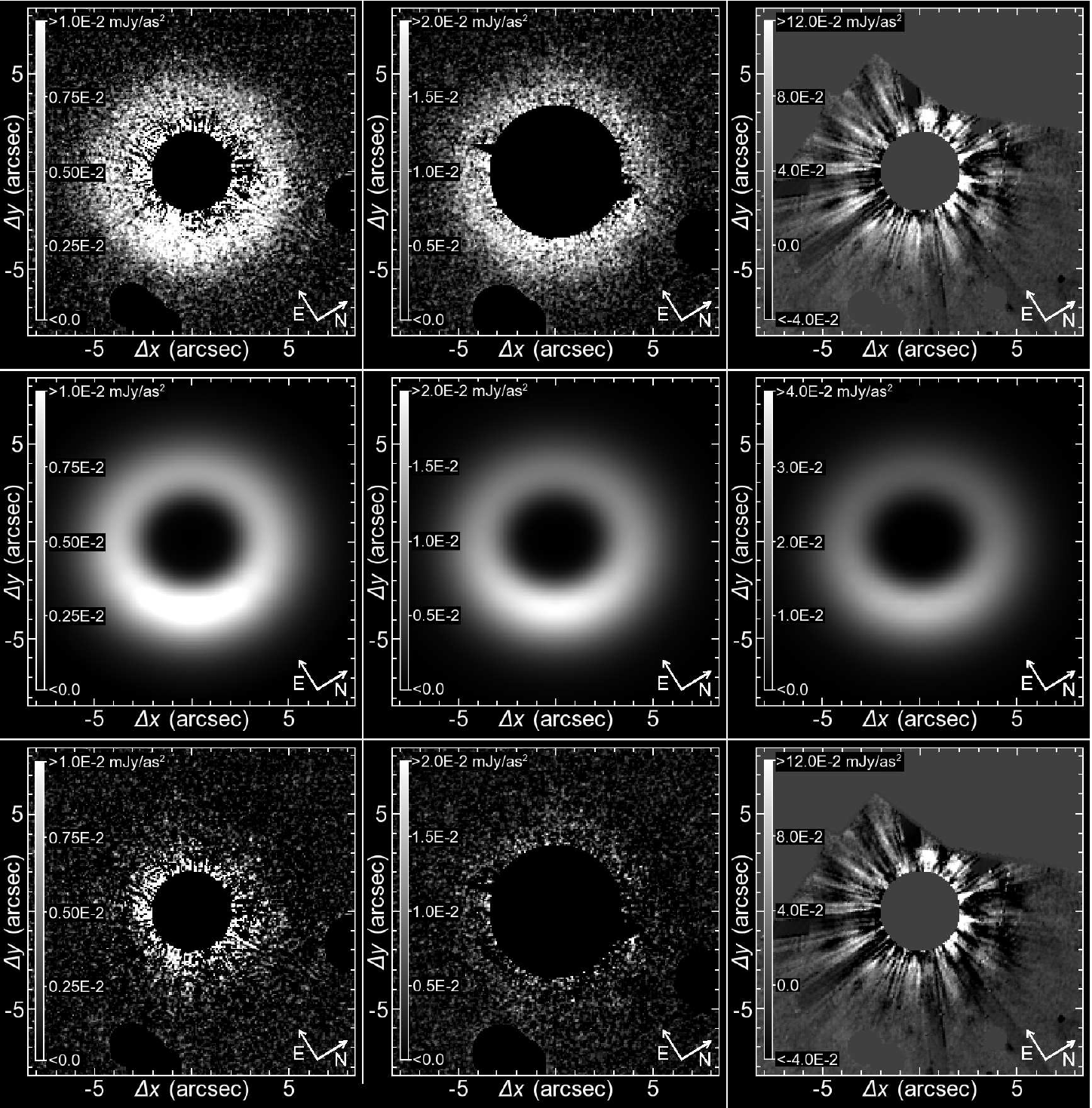}
\caption{Observed (top), modeled (center), and model subtracted images (bottom) in the F606W (left), the F814W (center), and the F110W (right) {\it HST} filters using the final model derived in this paper. The images have been rotated, so that the major axis of the disk is oriented along the x-axis. Only the considered parts of the images are shown, while regions of large residuals from PSF subtraction or contaminated by other objects on the line of sight are masked. In the F606W and F814W images the ring-like structure of the disk is well reproduced by the model, and only the residuals from PSF subtraction of the central star remain visible. In the residual F110W image parts of the disk remain visible in the residual image suggesting that scattering at this wavelength is underestimated in the model. Note the different flux scales in different columns. The model image in the right column is displayed in a different way than the data and model subtracted data to better visualize the disk.}
\label{ref_mod_diff}%
\end{figure*}

To search for the best-fit model, we simulate scattered light images for a grid of model parameters ($\alpha_1$, $r_{\rm{p}}$, and $s$) in our analytical density distribution. After a first, coarse search a grid with smaller spacings centered on the best fit value of the coarse search is employed for our final fitting. The considered range of parameters and the employed spacings are listed in Table~\ref{paratab}. Each simulated scattered light image is convolved with the corresponding PSF, derived with the Tiny Tim software\footnote{http://www.stecf.org/instruments/TinyTim/tinytimweb} and is scaled to minimize the $\chi^{2}$ derived from the observed and modeled image and the error map:
\begin{equation}
  \chi^{2} = \sum_{i}\left(\frac{\rm{observation} - \rm{model}}{\rm{uncertainty}}\right)^{2}
\end{equation}
In this equation the sum runs over all pixels of the used images that have not been masked (Fig.~\ref{ref_mod_diff}). The derived best-fit $\chi^{2}$ for each model is used to evaluate the agreement between observed and simulated image.

\begin{table}
\caption{Parameter space for modeling the {\it HST}/ACS F606W image (upper, middle) and the SED (lower).}
\label{paratab}      
\begin{center}
\begin{tabular}{cccc}        
\hline\hline                 
  Parameter & Grid & Range & Spacing \\
\hline                        
  $\alpha_1$ & coarse & 0.0 -- 14.0 & 0.5 \\
  $r_{\rm{p}}$ [AU] & & 112.0 -- 129.0 & 1.0 \\
  $s$ [AU] & & -10.0 -- 60.0 & 1.0 \\
\hline
  $\alpha_1$ & fine & 4.5 -- 10.0 & 0.1 \\
  $r_{\rm{p}}$ [AU] & & 118.2 -- 122.7 & 0.3 \\
  $s$ [AU] & & 10.0 -- 50.0 & 1.0 \\
\hline
  $a_{\rm{min}}$ [$\,\mu\rm{m}$] & -- & 0.5 -- 20 & 0.1 \\
  $\gamma$ & & 2.0 -- 5.0 & 0.1 \\
\hline                                   
\end{tabular}
\end{center}
\end{table}

Fitting our density distribution based on the F606W image, we find best-fit values of $\alpha_1 = 7.0^{+0.6}_{-0.8}$, $r_{\rm{p}} = 120.3^{+0.4}_{-0.5}\,\rm{AU}$, and $s = 25.0^{+8.0}_{-4.0}\,\rm{AU}$, where the uncertainties are given as $1\sigma$ confidence levels. These uncertainties represent only the formal errors from the fitting process. The distribution of the derived values of $\chi^{2}$ is plotted in Fig.~\ref{map_paraspace}. The plotted confidence levels are estimated by computing the values of $\chi^{2}$ on a three dimensional grid in the parameter space and by estimating the probability $p$ for each model j using the equation
\begin{equation}
  p = \exp\left[-\left(\chi^{2}_{\rm j} - \chi^{2}_{\rm{best}}\right)\right]
\end{equation}

Using the derived density distribution, one is able to reproduce the behaviour of the radial surface brightness profiles of the F606W image and outside $\approx 100\,\rm{AU}$ of the F814W image, not affected by the PSF subtraction errors described earlier in this section (Fig.~\ref{surf_bright_mod_ACS}).

\subsection{Deriving $a_{\rm{min}}$ and $\gamma$ from the SED}
\label{composition_fitting}

In the previous section, we derived the radial dust density distribution from the scattered light images. Under the conditions described in Sect.~\ref{HST_fitting} the results of this approach are independent from the grain properties such as lower grain size and size distribution. Employing this density distribution and assuming that it is the same for all grain sizes we can now simulate the SED of the debris disk to derive best-fit values for the parameters $a_{\rm{min}}$ and $\gamma$ of our grain size distribution. We start with the values assumed in Sect.~\ref{HST_fitting} and do not limit ourselves on a certain range of the parameters. However, since we do not expect values too far from what we assumed in Sect.~\ref{HST_fitting}, we effectively search the range of $a_{\rm{min}}$ and $\gamma$ listed in Table~\ref{paratab}. For each fit the corresponding disk mass is derived and compared with the mass derived from the F606W image using the same model parameters (dust composition and radial distribution).

We do not find any solution giving total disk masses from SED and scattered light that differ by less than a factor of 1.4, where the masses derived from scattered light are in any case larger than that derived from the SED. Taking into account that these mass estimates have been derived from observations which potentially trace different fractions of the dust distribution (e.g., grains of different size), and the uncertainties in flux scaling that arise from photometric calibration and stellar PSF subtraction, the resulting mass range is rather narrow and the derived masses are considered to be consistent. However, we are not able to reach our initial goal to derive \emph{one} mass value from all data sets to make it one final model parameter as it would have been from fully simultaneous modeling. This discrepancy is discussed in Sect.~\ref{masses}. For this reason we decide to fit the SED employing the dust density distribution derived from the scattered light data, but without the constraint that the derived masses should be identical. Instead, we derive independent disk masses from each data set for our final, unified model with the radial dust distribution derived from the F606W image and the dust composition derived from the SED of the system.

Furthermore, we are not able to reproduce the {\it Spitzer} spectrum and the rest of the SED data simultaneously (Fig.~\ref{SED_fit}). Thus the {\it Spitzer} spectrum is only considered including the synthetic photometry derived from it at 13 and $33\,\mu\rm{m}$ \citep{hil08}. The discrepancies with the shape of the spectrum are discussed in Sect.~\ref{spitzer-fit}.

Keeping in mind the above caveats we continue fitting the data with our approach described in Sect.~\ref{guideline}. Employing the radial dust density distribution derived in Sect.~\ref{HST_fitting} and the approach to fit the SED described above, the best-fit values for the parameters describing the dust properties are $a_{\rm{min}} = 2.5\,\mu\rm{m}$ and $\gamma = 3.6$ (Fig.~\ref{SED_fit}). The SED can be only poorly reproduced using $a_{\rm{min}} < 2.0\,\mu\rm{m}$ or $a_{\rm{min}} > 3.0\,\mu\rm{m}$ and any value of $\gamma$ as well as using $\gamma < 3.5$ or $\gamma > 3.7$ and any value of $a_{\rm{min}}$. These values are considered as limits of the fitted values. Adopting these results for the grain properties instead of the assumptions made in Sect.~\ref{HST_fitting} we were able to confirm the results on the density distribution derived there. The value of $\gamma = 3.6 \pm 0.1$ is consistent with that expected from an equilibrium collisional cascade \citep[$\gamma = 3.5$;][]{doh69}.

\subsection{Verifying the model using the {\it CARMA} 1.3\,mm map.}
\label{veri_carma}

The scattered light images do not allow to determine the dust distribution within the inner 2\arcsec\ (57\,AU) from the star. Furthermore, the spatial dust distribution seen in scattered light is dominated by the distribution of the smallest grains and can put only weak constraints on the distribution of grains with radii larger than $\sim 100\,\mu\rm{m}$. To verify the spatial dust distribution derived from the scattered light images one can use the {\it CARMA} 1.3\,mm map (Fig.~\ref{ref_diff_carma}). However, in this image most of the disk is detected with an average surface brightness of only $\approx 3\sigma$ while there are two peaks reaching $\approx 5\sigma$. The value of $\sigma$ has been derived in regions of the image where no signal from the disk is expected. Due to its low SNR and the coarse angular resolution relative to the {\it HST} images we do not use the 1.3\,mm map to fit model parameters. However we verify that our model is consistent with the {\it CARMA} observations.

Therefore, we simulate images at 1.3\,mm using the model derived above. We simulate observations of the best-fit model image using the actual u-v spacings from the {\it CARMA} observations. These model u-v data points were subtracted from the observations, and the residual image was reconstructed using the same image parameters as in \citet{cor09}. To correctly scale the modeled data to the observed source flux, we minimize the remaining flux in the reconstructed, model subtracted image.

From the reconstructed, model subtracted image (Fig.~\ref{ref_diff_carma}) we find that our model reproduces the global behaviour of the {\it CARMA} map, even though better results for slightly lower peak radii and/or a shallower decrease of the density distribution in the inner regions might be found. While a small amount of residual flux remains visible in the inner region of the disk, in the outer regions the disk flux is over-subtracted. However, since negative flux values with a ringlike shape around the disk are already visible in the original image and the flux remaining in the inner region of the disk hardly exceeds $3\sigma$, we rate these deficiencies as insignificant. We conclude that our density distribution is consistent with the {\it CARMA} data, keeping in mind the large uncertainties. Furthermore, the two peaks visible in this image are subtracted very well by our radially symmetrical image and peak residuals are around $3\sigma$. This implies that these structures may be noise, rather than real structures in the disk induced by a massive planet orbiting the star within the inner edge of the disk as suggested by \citet{cor09}.

At this part of the work we can already conclude that the small particles seen in scattered light and the larger particles traced by the CARMA data are cospatial within the errors of the available data. This is important, since it supports the assumption of our model, that particles of all sizes are distributed in the same way.

\subsection{Fitting the final model to the available data and deriving disk masses}
\label{finalfit}

\begin{figure}
\centering
\includegraphics[angle=270,width=1\linewidth]{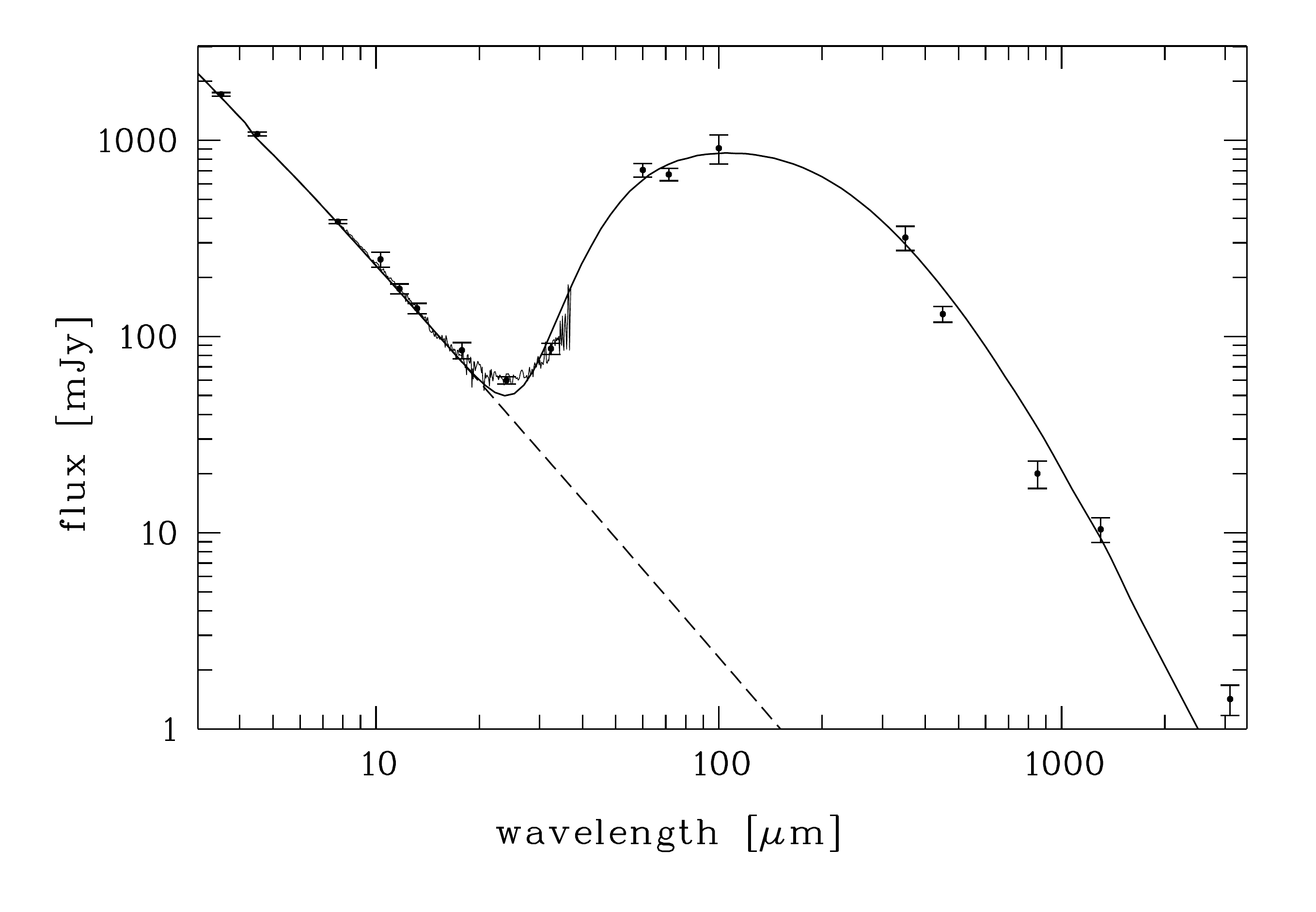}
\caption{Simulated SED from our single component best-fit model. The solid line represents the modeled SED, while the dashed line represents the stellar photospheric flux. Photometric measurements in the plotted wavelength range with corresponding error bars and the considered {\it Spitzer} spectrum are plotted, too. The discrepancies, in particular with the {\it Spitzer} spectrum are discussed in Sect.~\ref{spitzer-fit}.}
\label{SED_fit}%
\end{figure}

\begin{figure*}
\centering
\includegraphics[width=1\linewidth]{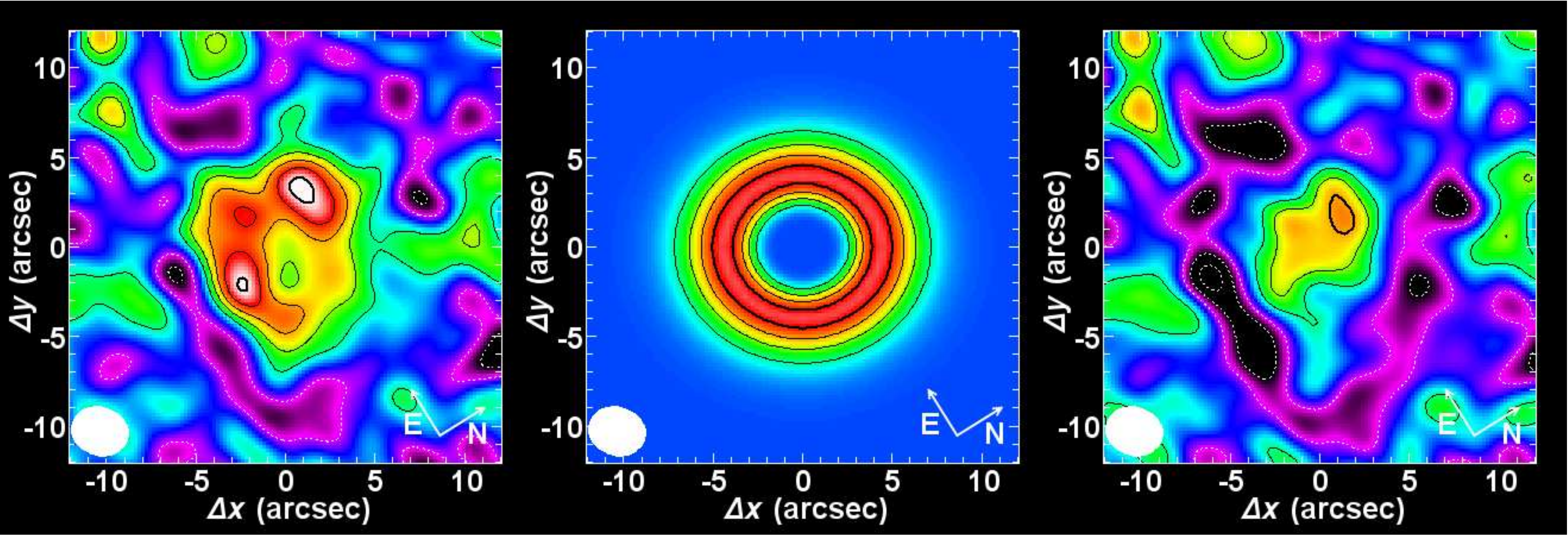}
\caption{Observed {\it CARMA} 1.3\,mm map (left), modeled image (center), and model subtracted map (right). The contour levels have increments of $1\sigma$. White, dashed levels start at $-1\sigma$ and indicate negative values, while black, solid levels start at $1\sigma$ and indicate positive values, where $1\sigma = 0.35\,\rm{mJy/beam}$. A bold line is used to mark the maximum contour in each image. The FWHM of the synthesized beam ($3\farcs2 \times 2\farcs7$) is shown in the lower-left corner. The images have been rotated, so that the major axis, as derived from the scattered light images, is oriented along the x-axis.}
\label{ref_diff_carma}%
\end{figure*}

In the above sections we employed the radial density distribution of the dust described in Sect.~\ref{density_dist} (Eq.~\ref{dens_final}) as well as the grain size distribution described there (Eq.~\ref{size_distr}) to derive a full set of model parameters ($\alpha_1$, $r_{\rm{p}}$, $s$, $a_{\rm{min}}$, and $\gamma$). We can now derive a total mass of the disk from each data set. This is done for the SED, the F606W, F814W, and F110W images by scaling the modeled data and subtracting them from the observations to minimize the residuals. We mask the inner 60\,AU in the F606W and F110W images, where the errors dominate significantly the signal. While the scattered light from the disk is subtracted very well in the F606W image, large parts of the disk remain visible after subtracting our best-fit model from the F110W image (Fig.~\ref{ref_mod_diff}). This seems to be due to a different distribution of the scattered light flux, closer to the star in this image, which results in a scaling factor for fitting the model to the image that is too small to reproduce the total flux. However, due to the large uncertainties in this image mentioned earlier, we do not further consider this fit. In the F814W image the inner 100\,AU affected by the large errors from PSF subtraction (Sect.~\ref{HST_fitting}) and two additional image artifacts with large negative values that affect significantly the fit are masked. The remaining disk structure can be reproduced very well (Fig.~\ref{surf_bright_mod_ACS},~\ref{ref_mod_diff}). The approach used for the CARMA data has been described in Sect.~\ref{veri_carma}. The slight overestimation of the disk extent compared to these data may explain the larger total disk mass derived from them compared to the mass derived from the SED.

The derived disk masses are listed in Table~\ref{modPara}. The scatter in the estimates based on the different data sets gives an idea of the self-consistency of our modeling approach with the data. Changing any of the model parameters within their confidence levels incurs correlated changes in some of the other parameters to maintain the quality of the fit (i.e., see Fig.~\ref{map_paraspace}), and affects the estimated disk masses by a factor of two to three. These changes in the disk mass are then correlated among the particular sets of data (scattered light images, SED and CARMA map).

\section{Discussion}
\label{discussion}

\subsection{Results}
\label{results}

We used one model (with the total disk mass derived from the different data sets varying by a factor of $\approx 1.5$) to reproduce the surface brightness in the F606W images and the outer parts of the F814W image being not affected by the PSF subtraction residuals discussed in Sect.~\ref{HST_fitting} (Fig.~\ref{ref_mod_diff},~\ref{surf_bright_mod_ACS}), the CARMA 1.3\,mm map (Fig.~\ref{ref_diff_carma}) and the SED of the HD\,107146 debris disk (Fig.~\ref{SED_fit}).

\begin{figure}
\centering
\includegraphics[angle=270,width=1\linewidth]{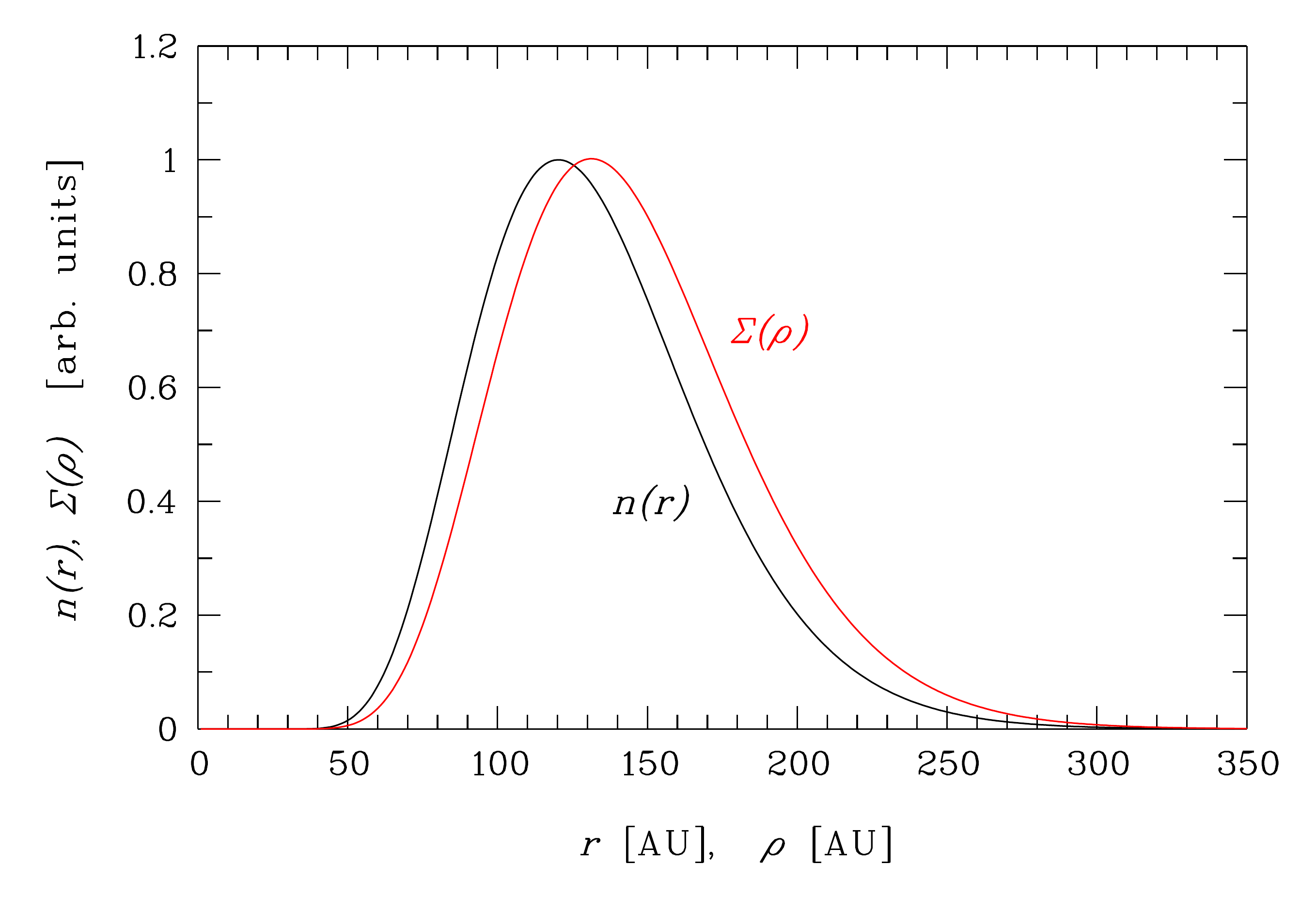}
\caption{Volume density distribution $n(r)$ over the radial distance $r$ from the star and surface density distribution $\Sigma(\rho)$ over the distance $\rho$ from the star projected on the disk midplane for our best-fit model (parameters see Table~\ref{modPara}). The analytical functions for $n(r)$ and $\Sigma(\rho)$ transform into each other following Eq.~\ref{eq1}. Both distributions are given in arbitrary units and scaled to a peak height of one for each distribution so that they can easily be compared. The peak particle density is $8.03 \times 10^{-6}\,{\rm m^{-3}}$, while the peak surface density amounts to $2.64 \times 10^{7}\,{\rm m^{-2}}$.}
\label{density}%
\end{figure}

We find the disk to be a broad ring with a peak of the density distribution $n(r)$ at $120.3^{+0.4}_{-0.5}\,\rm{AU}$, which corresponds to $131.4^{+0.5}_{-0.6}\,\rm{AU}$ for the surface density distribution $\Sigma(r)$ (Fig.~\ref{density}). The radial density distribution and surface density distribution transform into each other following Eq.~\ref{eq1}. Using the derived confidence levels of the parameters, we find the FWHM of the density distribution to be $86^{+9}_{-12}\,\rm{AU}$ which corresponds to an FWHM of the surface density distribution of $91^{+11}_{-12}\,\rm{AU}$. This is in good agreement with the values derived by \citet{ard04, ard05}. We find a lower dust grain size of $2.5^{+0.5}_{-0.5}\,\mu{\rm m}$, inconsistent with the blow-out size of the system of $0.5\,{\mu\rm m}$. The exponent of the grain size distribution of $\gamma = -3.6^{+0.1}_{-0.1}$ is consistent with the value expected from an equilibrium collisional cascade \citep{doh69}. Table~\ref{modPara} summarizes the derived model parameters. For the SED as well as for each resolved image we derive a total dust mass, presented in the table, too.

The available data is consistent with all dust in the system having the same radial distribution. This can be expected from a collision dominated disk \citep{wya05}. We find from integrating the synthetic stellar spectrum and the modeled SED of the disk a fractional luminosity $L_{\rm{dust}} / L_{\rm{star}} = 1.07 \times 10^{-3}$ which supports the scenario of a heavily collision dominated disk. However, there are some data available that are reproduced by our model only in an unsatisfactory way, in particular the shape of the {\it Spitzer}/IRS spectrum. These discrepancies are discussed in the following section.

\begin{table}
\caption{Derived Model Parameters}            
\label{modPara}      
\begin{center}                          
\begin{tabular}{ccc}
\hline\hline                 
Parameter & Derived Value & Uncertainty* \\    
\hline                        
   $\alpha_1$ & $7.0$ & 6.2 ... 7.6 \\      
   $r_{\rm{p}}$ [AU] & $120.3$ & $119.8$ ... $120.7$ \\
   $s$ [AU] & $25$ & $21.0$ ... $33.0$ \\
   $a_{\rm{min}}$ [$\mu\rm{m}$] & $2.5$ & $2.0$ ... $3.0$ \\
   $a_{\rm{max}}$ [$\mu\rm{m}$] & $1000.0$ & (fixed) \\
   $\gamma$ & $3.6$ & $3.5$ ... $3.7$ \\
   $M_{\rm{F606W}}$ [$\rm{M_{\sun}}$] & $6.2 \times 10^{-7}$ & -- \\
   $M_{\rm{F814W}}$ [$\rm{M_{\sun}}$] & $6.5 \times 10^{-7}$ & -- \\
   $M_{\rm{F110W}}$ [$\rm{M_{\sun}}$] & $8.5 \times 10^{-7}$ & -- \\
   $M_{\rm{CARMA}}$ [$\rm{M_{\sun}}$] & $6.7 \times 10^{-7}$ & -- \\
   $M_{\rm{SED}} $ [$\rm{M_{\sun}}$] & $4.4 \times 10^{-7}$ & -- \\
\hline                                   
\end{tabular}
\end{center}
\tablefoot{The uncertainties are given as $1\sigma$ confidence levels. For details see Sects.~\ref{HST_fitting}, \ref{composition_fitting}, and \ref{finalfit}.}
\end{table}

\subsection{Remaining discrepancies between our model and available data}

We are able to reproduce most of the available data with only one global model of the disk. However, there remain some discrepancies that are discussed in this section:

\begin{itemize}
 \item At wavelengths $\lambda < 40\,\mu\rm{m}$ the modeled SED poorly reproduces the observed one sampled by the {\it Spitzer} spectrum (Fig.~\ref{SED_fit}).
 \item We find a minimum grain size that is by a factor of 5 larger than the expected blow out size.
 \item Residual flux remains visible at peak level of $\approx 3\sigma$ in the inner region of the model-subtracted {\it CARMA} 1.3\,mm map, while a slight over-subtraction is visible in the outer region of the disk. This implies a slightly lower extent of the disk at this wavelength and results in too large an estimate of the total disk mass from this image.
 \item Significant parts of the disk remain visible in the F110W image after subtracting our best-fit model, while we already derived a mass that is by a factor of $\approx 1.3$ larger than that derived from the other scattered light images.
 \item We were not able to derive a unique value for the mass from all data sets. In contrast, the derived masses differ by a factor of up to 1.5.
\end{itemize}

While the model allows to reproduce the overall structure of the disk very well, important insights may be gained from the remaining discrepancies. They may be ascribed to a number of deficiencies of our model that become only relevant being faced with such a large number of high quality complementary data:
\begin{itemize}
 \item A discontinuity of the radial density distribution in the inner regions of the disk hidden by the coronagraph in the scattered light observations (e.g., an additional, inner disk component) would be able to significantly alter the shape of the SED at the short wavelength edge of the excess.
 \item The size distribution of the smallest grains may be only poorly described by a single power-law that describes the distribution of the larger grains very well. More detailed grain size distributions can be derived by detailed dynamical and collisional modeling of debris disks \citep[e.g.][]{the03, kri06, the07, kri08}. The results show a wavy distribution with an overdensity of small grains ($\approx$ twice the expected blow out size).
 \item A more complex chemical composition of the dust than the used astronomical silicate may be present. This has been studied for debris disks with significant emission features in the mid-infrared \citep[e.g.][]{bei05,lis07,lis08,che08}. A large variety of chemical dust compositions has been found, such as crystalline pyroxene, crystalline olivine, amorphous olivine, and water ice inclusions. For disks without such features the effect on the whole SED has been studied \citep[][Augereau et al.~2011 in preparation]{mue10}. Introducing porous grains instead of the compact, spherical grains used in this study also changes the optical properties of the dust \citep[e.g.,][]{vos06}.
 \item The derived density distribution as well as the surface brightness profiles derived from the scattered light data may be error-prone due to limitations from the PSF subtraction. This might result in errors for the derived radial density distribution of the dust and thus affect all derived model parameters.
 \item The spatial distribution in our model is the same for all grain sizes. In contrast, dynamical modeling of debris disks shows that several influences can result in a more complex dependence of the dust distribution on grain size \citep[e.g.][]{wya05, wya06}. The influences of radiation pressure and Poynting-Robertson drag, that increase with decreasing grain size (increasing $\beta$-ratio), may effect the distribution of the smallest grains which results in a segregation of grains of different sizes. Thus, a narrow ring of planetesimals can produce a broad disk through collisions \citep[birth ring scenario;][]{str06, kri06, kri10}. However, since we find the disk to be collision dominated, we do not expect grain segregation to be a dominant process for this disk \citep{wya05}. Even the presence of a massive planet does not result in strong radial grain segregation, although different spatial structures can be expected for different grain size \citep{wya06}.
\end{itemize}

In particular the effect of the first two points might have a strong effect on the shape of the SED at the short wavelength edge of the excess. Hence, we discuss possible extensions of our model in the following to account for these deficiencies. Exploring the chemical composition of the dust, in particular the shape of the dust grains, requires significant additional computational effort and is beyond the scope of the present paper.

\subsubsection{Unsatisfactory fit to the \emph{Spitzer}/IRS spectrum}
\label{spitzer-fit}

The most obvious and significant discrepancy of our model with the data is the different shape of the modeled SED in the range of the {\it Spitzer}/IRS spectrum. This might be explained in different ways. A more detailed grain size distribution than a power-law distribution as discussed above might result in a better fit to these data. Our model is able to account for such changes in the over all grain size distribution only by changing the power-law exponent and the lower grain size of the applied distribution, which might be insufficient. Since small (warm) grains dominate the emission at the short wavelength end of the excess, such an overabundance of small grains is expected to result in an increase of the simulated flux in the range of the \emph{Spitzer} spectrum. In such a scenario, one would expect that the additional excess flux seen in the \emph{Spitzer} spectrum originates from the same region of the disk like the long wavelength excess.

An alternative explanation for the discrepancy between our model and the \emph{Spitzer} spectrum is an additional, inner disk present in the system, such as discussed in the case of the $\epsilon$\,Eridani debris disk  \citep{bac09}. Such an additional, warm disk component would add excess flux at mid-infrared wavelengths and thus would be able to explain the missing flux from our model in this wavelength regime. In this case, the long wavelength and short wavelength excess should originate from different regions in the disk.

None of the available, resolved data allow us to distinguish between the two scenarios. The \emph{HST} data do not allow to constrain the dust distribution in the innermost regions of the disk due to the coronagraphic nature of these observations. The comparison of the model with the CARMA data gives a hint for additional emission in the inner region of the disk, but the sensitivity of these data is too low, so that the deviation between model and data is insignificant. Furthermore, the disk is unresolved by the \emph{Spitzer}/MIPS observations.

Our approach to distinguish between the two scenarios, is to explore what one would have to add to reproduce the \emph{Spitzer} spectrum using
\begin{itemize}
 \item[(a)] An additional grain size component to mimic a more complex grain size distribution. Here the dust is spatially distributed in the same way as the other dust (Sect.~\ref{HST_fitting}). The chemical composition of the dust is the same as before (astronomical silicate). We use only one distinct grain size here, so this is the only additional, free parameter.
 \item[(b)] An additional, inner disk component. We use a disk model that is described by a power-law radial density distribution $n(a) \propto a^{\alpha^{\rm{add}}}$, ranging from an inner radius $R_{\rm{in}}^{\rm{add}}$ to an outer radius $R_{\rm{out}}^{\rm{add}}$. The grain size distribution is described by $n(a) \propto a^{-\gamma^{\rm{add}}}$, ranging from a lower grain size $a_{\rm{min}}^{\rm{add}}$ to a fixed upper grain size of 1\,mm (the same parameterization as the grain size distribution in our original model). We use a fixed opening angle of $10^\circ$, so that the vertical disk structure is described in the same way as in Sect.~\ref{HST_fitting}. The same chemical composition of the dust as before (astronomical silicate) is employed. Since there is a total of 5 free parameters here and none of them can be fixed by simple physical assumptions, it is clear that this fit will be very degenerate. However, we will be able to evaluate, whether this scenario is able to reproduce the shape of the \emph{Spitzer} spectrum at all.
\end{itemize} 
 
First, we change the parameters of our (outer disk) model so that the simulated flux of the system is smaller than the measured flux in the \emph{Spitzer} spectrum, but is still consistent with the long wavelength SED observed. Since our radial density distribution derived from the scattered light data is still valid, we can only change the parameters of the grain size distribution. We find an increase of the lower grain size $a_{\rm{min}}$ from $2.5\,\mu\rm{m}$ to $3.5\,\mu\rm{m}$ without changing the exponent of the distribution, but increasing the total disk mass derived from fitting the SED data from $4.4 \times 10^{-7}\,\rm{M_{\sun}}$ to $4.8 \times 10^{-7}\,\rm{M_{\sun}}$ well suited to reach this goal. We then subtract the modeled SED of the system from the \emph{Spitzer} spectrum and fit the remaining excess flux with the two additional disk components described above. The exact choice of $a_{\rm{min}}$ is rather arbitrary, since we are not able to distinguish between flux from the original disk model (with changed dust parameters) and the hypothetic additional component. Thus, it is not possible to derive any formal uncertainties on these parameters. However, if we find an additional component that is able to reproduce the \emph{Spitzer}/IRS spectrum, then in this approach the value of $a_{\rm{min}} = 3.5\,\mu\rm{m}$ can be considered as a lower limit on the lower dust grain size of the original dust component. Smaller values would violate the constraint, that the excess from this component shall be smaller than the excess measured by \emph{Spitzer}/IRS. On the other hand, almost any value larger than $a_{\rm{min}} = 3.5\,\mu\rm{m}$ will be possible, since the missing flux at longer wavelengths can then be reproduced in a wide range by changing the parameters of the second, additional component.

We find from our approach (a), that it is not possible to fit the shape of the \emph{Spitzer} spectrum with an additional grain size component (with the radial density distribution derived in Sect.~\ref{HST_fitting}). Furthermore, this would add significant flux at longer wavelengths. In any case the dust is too cold (too far from the star) in this approach.

\begin{figure}
\centering
\includegraphics[angle=270,width=1\linewidth]{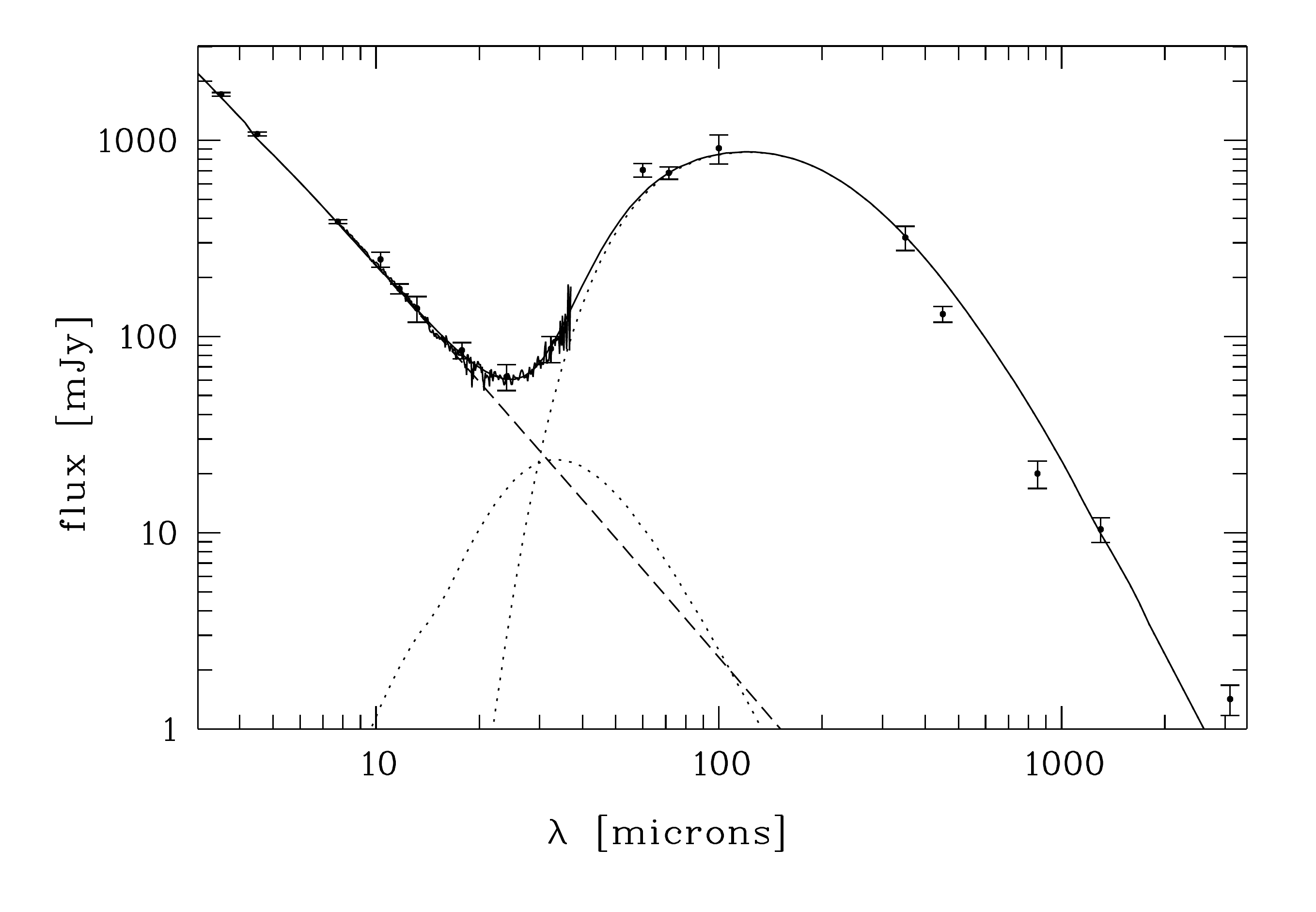}
\caption{SED fit with the modified dust composition in the outer disk and an additional, inner disk component (Sect.~\ref{spitzer-fit}). The two dotted curves indicate the emission from the two dust components only.}
\label{SED_new}%
\end{figure}

In contrast, with our approach (b) we are able to improve the fit on the \emph{Spitzer} spectrum significantly without lowering the quality of the fit on the other SED data. The explored parameter space for our search for an additional, inner disk component and our best-fit results are shown in Table~\ref{para_inner}. Fig.~\ref{SED_new} shows the new best-fit SED for this approach. Our best-fit gives a disk that is extended from the inner rim of the outer disk to very close to the star. The best-fit inner radius is identical with the smallest value of this parameter explored. This suggests that the disk is extended even to a lower inner radius. We find a nearly constant radial surface density distribution (exponent of radial density distribution $\alpha^{\rm{add}} = 0.8$ which corresponds to an exponent of the radial surface density distribution of -0.2) to be the best-fit. The lower dust grain size ($3.3\,\mu\rm{m}$) is very close to the lower grain size of the outer disk in this approach ($3.5\,\mu\rm{m}$). However, we find a very large exponent of the grain size distribution (best-fit: $\gamma^{\rm{add}} = 10$), i.e. a very narrow size distribution around the lowest grain size. The disk mass for this inner component then results to $3.6 \times 10^{-11}\,\rm{M}_{\sun}$. From this model we predict total scattered light fluxes from this disk component at $0.6\mu{\rm m}$ and $0.8\mu{\rm m}$ of 0.037\,mJy and 0.052\,mJy. The total reemitted flux from this component is plotted in Fig.~\ref{SED_new} as a function of the wavelength.

The spatial distribution of the dust as well as the steep grain size distribution and the lower grain size that is consistent with that in the outer disk suggest, that the inner disk component in this scenario is produced by collisions close to the inner rim of the outer disk and then dragged inwards towards the star by Poynting-Robertson drag. This scenario has proven successful to explain the origin of the dust in the habitable zone of $\epsilon$\,Eridani \citep{rei11}. For a transport dominated disk without perturbation by a planet, a constant radial surface density distribution has been found there from dynamical modeling. Such a disk should consist predominantly of small grains.

However, with the large uncertainties on our best-fit values it is not possible to draw strong conclusions from the resulting model. An additional, inner ring of planetesimals producing new dust through collisions (power-law exponent of the grain size distribution of 3.6) would also be possible within the $1\sigma$ uncertainties of our fit. It is important to note that these uncertainties are only the formal uncertainties of our fit on the remaining flux in the \emph{Spitzer} spectrum after subtraction of the SED from a suitable model of the outer disk. They do not include the degrees of freedom introduced by the selection of the parameters of the outer disk which are expected to dominate the real uncertainties on the fitted model parameters. For example, a further increase of the lower dust grain size in the outer disk would be possible and would make a less steep grain size distribution more probable.

\begin{table}
\caption{Parameter space explored and fitting results for modeling the remaining \emph{Spitzer}/IRS flux by an additional, inner disk component.}
\label{para_inner}      
\begin{center}
\begin{tabular}{ccccc}        
\hline\hline                 
  Parameter & Range & Spacing & Best-fit & Uncertainty$^{1}$\\
\hline                        
  $R_{\rm{in}}^{\rm{add}}$ [AU] & 0.2 -- 250.0 & temp$^{2}$ & 0.2 & 0.2 ... 0.6 \\
  $R_{\rm{out}}^{\rm{add}}$ [AU] & 2.0 -- 250.0 & temp$^{2}$ & 42.2 & 6.0 ... 250 \\
  $\alpha^{\rm{add}}$ & -5.0 -- 5.0 & 0.1 & 0.8 & -2.1 ... 1.1 \\
\hline
  $a_{\rm{min}}^{\rm{add}}$ [$\mu$m] & 0.5 -- 100 & log${^3}$ & 3.3 & 0.8 ... 4.4 \\
  $\gamma^{\rm{add}}$ & 2.0 -- 10.0 & 0.1 & 10.0 & 3.6 ... 10.0 \\
  $M^{\rm{add}}$ [$\rm{M}_{\sun}$]& free & cont$^{4}$ & $3.6 \times 10^{-11}$ & \dots \\
\hline                                   
\end{tabular}
\end{center}
\tablefoot{(1) Confidence levels ($1\sigma$); (2) equally distributed in temperature of the dust species with the largest gradient of the radial temperature distribution (spacing $= 2\,\rm{K}$); (3) equally distributed in $\log(a)$ (100 size bins); (4) continuous -- The fitting of this parameter has not been done on a grid, but a continuous range of possible values has been used.}
\end{table}

\subsubsection{Different disk masses derived from different data sets}
\label{masses}

While we were able to reproduce most of the available data using one model of the disk, we were not able to derive a unique disk mass. The derived masses from the different data sets differ by a factor of up to 1.5 (ignoring the {\it HST}/NICMOS data which show a larger deviation). In fully simultaneous modeling of all data, this mass would have been a parameter of the model that would have been to be fitted on the whole data set. This would have resulted in unsatisfactory fits on nearly all data sets, because of inadequate flux scaling of the different simulated data sets. The discrepancy is rather low, in particular keeping in mind the photometric uncertainties in the single measurements, uncertainties from imperfect PSF subtraction, and the fact that we compare values derived from scattered light data and thermal reemission data. However, a global model of the disk should be able to reproduce all data without the need of different flux scaling done by deriving different masses from different data sets.

Again, a more complex grain size distribution than the expected one is one possible explanation. A different slope of the grain size distribution for the smallest grains present in the system would result in a different surface brightness at scattered light wavelengths, since these observations are very sensitive to the smallest grains. The slope of the grain size distribution in our single power-law model is mostly constrained by the shape of the SED at longer wavelengths ($\lambda \geq 100\,\mu\rm{m}$, which is most sensitive to larger grains. A derived disk mass from scattered light data that is too large (as it is the case for our results) would mean a too low surface brightness of our modeled data at these wavelengths. An underabundance of the smallest grains in our model would be a plausible explanation. However, we found in Sect.~\ref{spitzer-fit}, that adding an additional grain size component is not able to solve the discrepancy with the \emph{Spitzer} spectrum. Furthermore, the smallest grain size is well constrained by the short wavelength end of the excess for the adopted chemical dust composition. Even with an additional inner disk component (Sect.~\ref{spitzer-fit}) there is a strong lower limit for the smallest grains. Thus, a different chemical dust composition, e.g., with larger albedo to increase the scattering efficiency of the dust, is the most plausible explanation.

An alternative to a different chemical composition of the dust could be a different structure of the grains, i.e. the presence of fluffy, porous grains \citep[e.g.,][]{vos06, vos08}. This might alter the absorption efficiency and scattering efficiency and the wavelength dependence of these quantities independently.

\subsubsection{A large lower dust grain size}
\label{lowest_grains}

The lower grain size from dynamical and collisional models of debris disks is in any case consistent with the expected blow out size, while there is an overdensity of particles with $\approx$ twice the blow out size \citep[e.g.][]{the03}. In contrast, we find a lower grain size of $\approx 5$ times the expected blow out size for our original model and even larger for the model with the additional inner disk component. Because small grains are in general hotter than large ones, they are traced particularly by the short wavelength end of the IR excess SED. While the lower grain size $a_{\rm{min}}$ is usually heavily degenerate with the spatial dust distribution in pure SED modeling, this is well constrained in the present work by the scattered light images. As the smallest grains are the most efficiently scattering ones, our radial density distribution derived from scattered light images is expected to be particularly reliable for these grains. Thus, grain segregation is not expected to have a significant effect on the lower grain size in our modeling approach. To lower the value of $a_{\rm{min}}$ in our derived solution one can solely change the chemical composition or shape of the dust grains to get in general cooler dust. The inclusion of water ice has been successful to solve this problem (Augereau et al. 2011, in preparation). As discussed before, adding smaller grains would in general also increase the scattering efficiency of the dust, increasing the modeled surface brightness and thus decreasing the mass derived from the scattered light images to a level more consistent with the value derived from fitting the SED. A different chemical dust composition or shape might also increase the value of $\beta$ of the dust and thus increase the expected blow-out size to a level more consistent with the findings of our modeling. Again, a different structure of the grains might have a very similar effect.

\subsection{Is the HD\,107146 debris disk unusually massive for its age?}

Compared to other stars harboring debris disks, HD\,107146 is in an intermediate age of $\approx 100\,\rm{Myr}$ to $150\,\rm{Myr}$. It is harboring a very massive debris disk. Such high dust masses are much more usual for young debris disks around A type stars. Spatially resolved debris disks with comparable dust masses are for example $\beta$\,Pic \citep{zuc93}, HD\,99803B \citep{wya03a}, HR\,4796 \citep{she04}, and HR\,8799 \citep{su09}. All these stars are A type stars or earlier. Of these stars HD99803B and HR\,8799 might have a comparable age to HD\,107146, while the two other ones are significantly younger. The dust mass in debris disks is expected to decline with age trough well explored depletion mechanisms \citep[e.g.][]{sil00, loe08, roc09} and towards later spectral types due to lower initial disk mass. In contrast, HD\,107146 is an intermediate age G type star harboring a disk with a mass comparable to the mass of disks around young A-type stars. This suggests that the HD\,107146 disk is special among the known, spatially resolved debris disks. Its high age and high dust mass may imply that the system is undergoing an analog to the Late Heavy Bombardment \citep[delayed stirring;][]{dom03} in our own solar system as also suggested for HR\,8799 \citep{su09}.

\subsection{The origin of the disk shape}

The shape of the HD\,107146 debris disk is similar to that of Fomalhaut (HD\,216956), an A3V star of 200\,Myr \citep{hab01}, although with $6.2 \times 10^{-8}\,\rm{M_{\sun}}$ \citep{she04} Fomalhaut's debris disk has a lower mass by about one order of magnitude. Both stars have comparable ages. The similar shape of the two disks and the fact that the shape of the Fomalhaut disk seems to be induced by the planet Fomalhaut~b orbiting the Star \citep{kal08} raises the question if a planet is embedded also in the disk of HD\,107146. \citet{ken04} found from numerical simulations, that bright ring-like structures in debris disks are a by-product of planet formation while \citet{the07} simulated the collisional evolution of debris disks without perturbing planets and without initial ringlike structure (e.g. a birth ring scenario) and found no evidence for the formation of ring-like structures or inner gaps. However, while the inner edge of Fomalhaut's debris disk represents a sharp break in particle density, the inner edge of the HD\,107146 debris disk seems to be a shallow decrease. Furthermore, the Fomalhaut disk is off-centered from the stellar position which is interpreted as a result of the planet-disk interaction. There is no obvious evidence in the data that the HD\,107146 disk is off-centered in a similar way. Alternatively, a birth ring scenario of a collision dominated disk (as it is expected for the HD\,107146 debris disk) is also able to explain inner gaps in debris disks without the need a massive interior planet \citep{kri10}. Here the planetesimals do not necessarily have to be distributed in a narrow ring, but the radial broadness of the planetesimal belt has to be similar to or smaller than that of the observed dust disk. Furthermore, the peak surface density of the planetesimal belt is expected to be close to the peak position of the surface density of the dust disk ($\approx 130\,{\rm AU}$ for HD\,107146).

Further {\it HST} coronagraphic observations have been approved for execution to be conducted as part of the Cycle 18 program GO 12228\footnote{http://www.stsci.edu/cgi-bin/get-proposal-info?12228} (G. Schneider, PI), where 8 orbits of STIS coronagraphy in the broadband optical are devoted to probe the inner regions of the HD\,107146 disk as close as $r = 0.3\arcsec$ (8.6\,AU) and to look for resonant structures in the dust distribution. These data may, through higher quality imaging, better constrain possible disk sub structures induced by planets as well as on the reliability of our derived density distribution, particularly in the inner regions of the disk.

\section{Conclusions}
\label{conc}

A detailed model of the debris disk around HD\,107146 has been created and fitted to all available data using the approach described in the paper. A broad range of resolved data from optical scattered light to millimeter data as well as the SED has been reproduced by a single radial density profile. All parameters of our model have been fixed without remaining degeneracies using only one scattered light image and the well-sampled SED of the disk in the particular case (Sect.~\ref{modeling}). We found the disk to be a broad ring of dust with smooth inner and outer boundaries, and with a peak position from the star and FWHM of the surface density of $\approx 131\,\rm{AU}$ and $\approx 91\,\rm{AU}$, respectively. We concluded from our modeling results that the disk is heavily collision dominated. Discrepancies of our model to the data are discussed. From these discrepancies we find strong evidence for an additional inner disk component, possibly near the habitable zone (the stellocentric distance where liquid water can exist on planetary surfaces, i.e., few AU from a solar-type star) of the star and for a different chemical composition or shape of the dust grains. The additional inner disk component is predominantly composed of small grains and the results from our modeling are consistent with these grains being released at the inner edge of the disk and dragged inwards due to Poynting-Robertson drag. Only detailed, simultaneous modeling of the large amount of available high quality, complementary data was able to reveal these signposts. We did not find evidence of an orbiting planet from the available data and found that a birth ring scenario is likely responsible for the ringlike shape of the disk.

\begin{acknowledgements}
This work is based in part on observations made with the NASA/ESA {\it Hubble Space Telescope}, obtained at the Space Telescope Science Institute, which is operated by the Association of Universities for Research in Astronomy, Inc., under NASA contract NAS 5-26555. These observations are associated with program \# 10177. Support for program \# 10177 was provided by NASA through a grant from the Space Telescope science Institute. This work is also based in part on observations made with the {\it Spitzer Space Telescope}, which is operated by JPL/Caltech under NASA contract 1407. These observations are associated with the Spitzer Legacy Science Program FEPS, which is supported through NASA contracts 1224768, 1224634, and 1224566 administered through JPL. We thank all members of the FEPS team for their contributions to this effort. We would also like to particularly thank David Ardila for providing the {\it HST}/ACS data which represent the basis for the recent work, and Andrea Isella for producing the model subtracted images of the {\it CARMA} data. S. E. thanks for financial support from DFG under contract WO\,857/7-1.
\end{acknowledgements}

\bibliographystyle{aa}

\bibliography{bibtex}

\begin{thebibliography}{56}
\expandafter\ifx\csname natexlab\endcsname\relax\def\natexlab#1{#1}\fi

\bibitem[{{Ardila} {et~al.}(2004){Ardila}, {Golimowski}, {Krist}, {Clampin},
  {Williams}, {Blakeslee}, {Ford}, {Hartig}, \& {Illingworth}}]{ard04}
{Ardila}, D.~R., {Golimowski}, D.~A., {Krist}, J.~E., {et~al.} 2004, \apjl,
  617, L147

\bibitem[{{Ardila} {et~al.}(2005){Ardila}, {Golimowski}, {Krist}, {Clampin},
  {Williams}, {Blakeslee}, {Ford}, {Hartig}, \& {Illingworth}}]{ard05}
{Ardila}, D.~R., {Golimowski}, D.~A., {Krist}, J.~E., {et~al.} 2005, \apjl,
  624, L141

\bibitem[{{Aumann} {et~al.}(1984){Aumann}, {Beichman}, {Gillett}, {de Jong},
  {Houck}, {Low}, {Neugebauer}, {Walker}, \& {Wesselius}}]{aum84}
{Aumann}, H.~H., {Beichman}, C.~A., {Gillett}, F.~C., {et~al.} 1984, \apjl,
  278, L23

\bibitem[{{Backman} {et~al.}(2009){Backman}, {Marengo}, {Stapelfeldt}, {Su},
  {Wilner}, {Dowell}, {Watson}, {Stansberry}, {Rieke}, {Megeath}, {Fazio}, \&
  {Werner}}]{bac09}
{Backman}, D., {Marengo}, M., {Stapelfeldt}, K., {et~al.} 2009, \apj, 690, 1522

\bibitem[{{Beichman} {et~al.}(2005){Beichman}, {Bryden}, {Gautier},
  {Stapelfeldt}, {Werner}, {Misselt}, {Rieke}, {Stansberry}, \&
  {Trilling}}]{bei05}
{Beichman}, C.~A., {Bryden}, G., {Gautier}, T.~N., {et~al.} 2005, \apj, 626,
  1061

\bibitem[{{Boffi} {et~al.}(2007)}]{bof07}
{Boffi}, F.~R. {et~al.} 2007, Baltimore: STScI

\bibitem[{{Carpenter} {et~al.}(2005){Carpenter}, {Wolf}, {Schreyer},
  {Launhardt}, \& {Henning}}]{car05}
{Carpenter}, J.~M., {Wolf}, S., {Schreyer}, K., {Launhardt}, R., \& {Henning},
  T. 2005, \aj, 129, 1049

\bibitem[{{Chen} {et~al.}(2008){Chen}, {Fitzgerald}, \& {Smith}}]{che08}
{Chen}, C.~H., {Fitzgerald}, M.~P., \& {Smith}, P.~S. 2008, \apj, 689, 539

\bibitem[{{Corder} {et~al.}(2009){Corder}, {Carpenter}, {Sargent}, {Zauderer},
  {Wright}, {White}, {Woody}, {Teuben}, {Scott}, {Pound}, {Plambeck}, {Lamb},
  {Koda}, {Hodges}, {Hawkins}, \& {Bock}}]{cor09}
{Corder}, S., {Carpenter}, J.~M., {Sargent}, A.~I., {et~al.} 2009, \apjl, 690,
  L65

\bibitem[{{Dohnanyi}(1969)}]{doh69}
{Dohnanyi}, J.~S. 1969, \jgr, 74, 2531

\bibitem[{{Dominik} \& {Decin}(2003)}]{dom03}
{Dominik}, C. \& {Decin}, G. 2003, \apj, 598, 626

\bibitem[{{Draine} \& {Lee}(1984)}]{dra84}
{Draine}, B.~T. \& {Lee}, H.~M. 1984, \apj, 285, 89

\bibitem[{{Habing} {et~al.}(2001){Habing}, {Dominik}, {Jourdain de Muizon},
  {Laureijs}, {Kessler}, {Leech}, {Metcalfe}, {Salama}, {Siebenmorgen},
  {Trams}, \& {Bouchet}}]{hab01}
{Habing}, H.~J., {Dominik}, C., {Jourdain de Muizon}, M., {et~al.} 2001, \aap,
  365, 545

\bibitem[{{Hillenbrand} {et~al.}(2008){Hillenbrand}, {Carpenter}, {Kim},
  {Meyer}, {Backman}, {Moro-Mart{\'{\i}}n}, {Hollenbach}, {Hines}, {Pascucci},
  \& {Bouwman}}]{hil08}
{Hillenbrand}, L.~A., {Carpenter}, J.~M., {Kim}, J.~S., {et~al.} 2008, \apj,
  677, 630

\bibitem[{{Kalas} {et~al.}(2008){Kalas}, {Graham}, {Chiang}, {Fitzgerald},
  {Clampin}, {Kite}, {Stapelfeldt}, {Marois}, \& {Krist}}]{kal08}
{Kalas}, P., {Graham}, J.~R., {Chiang}, E., {et~al.} 2008, Science, 322, 1345

\bibitem[{{Kalas} {et~al.}(2006){Kalas}, {Graham}, {Clampin}, \&
  {Fitzgerald}}]{kal06}
{Kalas}, P., {Graham}, J.~R., {Clampin}, M.~C., \& {Fitzgerald}, M.~P. 2006,
  \apjl, 637, L57

\bibitem[{{Kenyon} \& {Bromley}(2004)}]{ken04}
{Kenyon}, S.~J. \& {Bromley}, B.~C. 2004, \aj, 127, 513

\bibitem[{{Krivov}(2010)}]{kri10}
{Krivov}, A.~V. 2010, Research in Astronomy and Astrophysics, 10, 383

\bibitem[{{Krivov} {et~al.}(2006){Krivov}, {L{\"o}hne}, \& {Srem{\v
  c}evi{\'c}}}]{kri06}
{Krivov}, A.~V., {L{\"o}hne}, T., \& {Srem{\v c}evi{\'c}}, M. 2006, \aap, 455,
  509

\bibitem[{{Krivov} {et~al.}(2008){Krivov}, {M{\"u}ller}, {L{\"o}hne}, \&
  {Mutschke}}]{kri08}
{Krivov}, A.~V., {M{\"u}ller}, S., {L{\"o}hne}, T., \& {Mutschke}, H. 2008,
  \apj, 687, 608

\bibitem[{{Kurucz}(1969)}]{kur69}
{Kurucz}, R.~L. 1969, \apj, 156, 235

\bibitem[{{Lisse} {et~al.}(2007){Lisse}, {Beichman}, {Bryden}, \&
  {Wyatt}}]{lis07}
{Lisse}, C.~M., {Beichman}, C.~A., {Bryden}, G., \& {Wyatt}, M.~C. 2007, \apj,
  658, 584

\bibitem[{{Lisse} {et~al.}(2008){Lisse}, {Chen}, {Wyatt}, \& {Morlok}}]{lis08}
{Lisse}, C.~M., {Chen}, C.~H., {Wyatt}, M.~C., \& {Morlok}, A. 2008, \apj, 673,
  1106

\bibitem[{{L{\"o}hne} {et~al.}(2008){L{\"o}hne}, {Krivov}, \&
  {Rodmann}}]{loe08}
{L{\"o}hne}, T., {Krivov}, A.~V., \& {Rodmann}, J. 2008, \apj, 673, 1123

\bibitem[{{Maybhate} {et~al.}(2010){Maybhate}, {Armstrong}, {et~al.}}]{may10}
{Maybhate}, A., {Armstrong}, A., {et~al.} 2010, Baltimore: STScI

\bibitem[{{Metchev} {et~al.}(2004){Metchev}, {Hillenbrand}, \& {Meyer}}]{met04}
{Metchev}, S.~A., {Hillenbrand}, L.~A., \& {Meyer}, M.~R. 2004, \apj, 600, 435

\bibitem[{{Meyer} {et~al.}(2007){Meyer}, {Backman}, {Weinberger}, \&
  {Wyatt}}]{mey07}
{Meyer}, M.~R., {Backman}, D.~E., {Weinberger}, A.~J., \& {Wyatt}, M.~C. 2007,
  Protostars and Planets V, 573

\bibitem[{{Meyer} {et~al.}(2006){Meyer}, {Hillenbrand}, {Backman}, {Beckwith},
  {Bouwman}, {Brooke}, {Carpenter}, {Cohen}, {Cortes}, {Crockett}, {Gorti},
  {Henning}, {Hines}, {Hollenbach}, {Kim}, {Lunine}, {Malhotra}, {Mamajek},
  {Metchev}, {Moro-Martin}, {Morris}, {Najita}, {Padgett}, {Pascucci},
  {Rodmann}, {Schlingman}, {Silverstone}, {Soderblom}, {Stauffer}, {Stobie},
  {Strom}, {Watson}, {Weidenschilling}, {Wolf}, \& {Young}}]{mey06}
{Meyer}, M.~R., {Hillenbrand}, L.~A., {Backman}, D., {et~al.} 2006, \pasp, 118,
  1690

\bibitem[{{Mo{\'o}r} {et~al.}(2006){Mo{\'o}r}, {{\'A}brah{\'a}m}, {Derekas},
  {Kiss}, {Kiss}, {Apai}, {Grady}, \& {Henning}}]{moo06}
{Mo{\'o}r}, A., {{\'A}brah{\'a}m}, P., {Derekas}, A., {et~al.} 2006, \apj, 644,
  525

\bibitem[{{M{\"u}ller} {et~al.}(2010){M{\"u}ller}, {L{\"o}hne}, \&
  {Krivov}}]{mue10}
{M{\"u}ller}, S., {L{\"o}hne}, T., \& {Krivov}, A.~V. 2010, \apj, 708, 1728

\bibitem[{{Najita} \& {Williams}(2005)}]{naj05}
{Najita}, J. \& {Williams}, J.~P. 2005, \apj, 635, 625

\bibitem[{{Perryman} {et~al.}(1997){Perryman}, {Lindegren}, {Kovalevsky},
  {Hoeg}, {Bastian}, {Bernacca}, {Cr{\'e}z{\'e}}, {Donati}, {Grenon}, {van
  Leeuwen}, {van der Marel}, {Mignard}, {Murray}, {Le Poole}, {Schrijver},
  {Turon}, {Arenou}, {Froeschl{\'e}}, \& {Petersen}}]{per97}
{Perryman}, M.~A.~C., {Lindegren}, L., {Kovalevsky}, J., {et~al.} 1997, \aap,
  323, L49

\bibitem[{{Reidemeister} {et~al.}(2011){Reidemeister}, {Krivov}, {Stark},
  {Augereau}, {L{\"o}hne}, \& {M{\"u}ller}}]{rei11}
{Reidemeister}, M., {Krivov}, A.~V., {Stark}, C.~C., {et~al.} 2011, \aap, 527,
  A57+

\bibitem[{{Roccatagliata} {et~al.}(2009){Roccatagliata}, {Henning}, {Wolf},
  {Rodmann}, {Corder}, {Carpenter}, {Meyer}, \& {Dowell}}]{roc09}
{Roccatagliata}, V., {Henning}, T., {Wolf}, S., {et~al.} 2009, \aap, 497, 409

\bibitem[{{Schneider} {et~al.}(2005){Schneider}, {Silverstone}, \&
  {Hines}}]{schn05}
{Schneider}, G., {Silverstone}, M.~D., \& {Hines}, D.~C. 2005, \apjl, 629, L117

\bibitem[{{Schneider} {et~al.}(2006){Schneider}, {Silverstone}, {Hines},
  {Augereau}, {Pinte}, {M{\'e}nard}, {Krist}, {Clampin}, {Grady}, {Golimowski},
  {Ardila}, {Henning}, {Wolf}, \& {Rodmann}}]{schn06}
{Schneider}, G., {Silverstone}, M.~D., {Hines}, D.~C., {et~al.} 2006, \apj,
  650, 414

\bibitem[{{Sheret} {et~al.}(2004){Sheret}, {Dent}, \& {Wyatt}}]{she04}
{Sheret}, I., {Dent}, W.~R.~F., \& {Wyatt}, M.~C. 2004, \mnras, 348, 1282

\bibitem[{{Silverstone}(2000)}]{sil00}
{Silverstone}, M.~D. 2000, PhD thesis, UNIVERSITY OF CALIFORNIA, LOS ANGELES

\bibitem[{{Smith} \& {Terrile}(1984)}]{smi84}
{Smith}, B.~A. \& {Terrile}, R.~J. 1984, Science, 226, 1421

\bibitem[{{Strubbe} \& {Chiang}(2006)}]{str06}
{Strubbe}, L.~E. \& {Chiang}, E.~I. 2006, \apj, 648, 652

\bibitem[{{Su} {et~al.}(2009){Su}, {Rieke}, {Stapelfeldt}, {Malhotra},
  {Bryden}, {Smith}, {Misselt}, {Moro-Martin}, \& {Williams}}]{su09}
{Su}, K.~Y.~L., {Rieke}, G.~H., {Stapelfeldt}, K.~R., {et~al.} 2009, \apj, 705,
  314

\bibitem[{{Th{\'e}bault} \& {Augereau}(2007)}]{the07}
{Th{\'e}bault}, P. \& {Augereau}, J. 2007, \aap, 472, 169

\bibitem[{{Th{\'e}bault} {et~al.}(2003){Th{\'e}bault}, {Augereau}, \&
  {Beust}}]{the03}
{Th{\'e}bault}, P., {Augereau}, J.~C., \& {Beust}, H. 2003, \aap, 408, 775

\bibitem[{{Voshchinnikov} \& {Henning}(2008)}]{vos08}
{Voshchinnikov}, N.~V. \& {Henning}, T. 2008, \aap, 483, L9

\bibitem[{{Voshchinnikov} {et~al.}(2006){Voshchinnikov}, {Il'in}, {Henning}, \&
  {Dubkova}}]{vos06}
{Voshchinnikov}, N.~V., {Il'in}, V.~B., {Henning}, T., \& {Dubkova}, D.~N.
  2006, \aap, 445, 167

\bibitem[{{Weinberger} {et~al.}(1999){Weinberger}, {Becklin}, {Schneider},
  {Smith}, {Lowrance}, {Silverstone}, {Zuckerman}, \& {Terrile}}]{wei99}
{Weinberger}, A.~J., {Becklin}, E.~E., {Schneider}, G., {et~al.} 1999, \apjl,
  525, L53

\bibitem[{{Weingartner} \& {Draine}(2001)}]{wei01}
{Weingartner}, J.~C. \& {Draine}, B.~T. 2001, \apj, 548, 296

\bibitem[{{Williams} {et~al.}(2004){Williams}, {Najita}, {Liu}, {Bottinelli},
  {Carpenter}, {Hillenbrand}, {Meyer}, \& {Soderblom}}]{wil04}
{Williams}, J.~P., {Najita}, J., {Liu}, M.~C., {et~al.} 2004, \apj, 604, 414

\bibitem[{{Wolf} \& {Hillenbrand}(2003)}]{wol03}
{Wolf}, S. \& {Hillenbrand}, L.~A. 2003, \apj, 596, 603

\bibitem[{{Wolf} \& {Hillenbrand}(2005)}]{wol05}
{Wolf}, S. \& {Hillenbrand}, L.~A. 2005, Computer Physics Communications, 171,
  208

\bibitem[{{Wolf} \& {Voshchinnikov}(2004)}]{wol04}
{Wolf}, S. \& {Voshchinnikov}, N.~V. 2004, Computer Physics Communications,
  162, 113

\bibitem[{{Wyatt}(2005)}]{wya05}
{Wyatt}, M.~C. 2005, \aap, 433, 1007

\bibitem[{{Wyatt}(2006)}]{wya06}
{Wyatt}, M.~C. 2006, \apj, 639, 1153

\bibitem[{{Wyatt}(2008)}]{wya08}
{Wyatt}, M.~C. 2008, \araa, 46, 339

\bibitem[{{Wyatt} {et~al.}(2003){Wyatt}, {Dent}, \& {Greaves}}]{wya03a}
{Wyatt}, M.~C., {Dent}, W.~R.~F., \& {Greaves}, J.~S. 2003, \mnras, 342, 876

\bibitem[{{Zuckerman} \& {Becklin}(1993)}]{zuc93}
{Zuckerman}, B. \& {Becklin}, E.~E. 1993, \apj, 414, 793

\end{thebibliography}

\end{document}